\documentclass[twocolumn]{aastex631}

\usepackage{style}

\begin{document}

\title{The NANOGrav 15 yr Data Set: Piecewise Power-Law\\ Reconstruction of the Gravitational-Wave Background}
\author[0000-0001-5134-3925]{Gabriella Agazie}
\affiliation{Center for Gravitation, Cosmology and Astrophysics, Department of Physics and Astronomy, University of Wisconsin-Milwaukee,\\ P.O. Box 413, Milwaukee, WI 53201, USA}
\email{gabriella.agazie@nanograv.org}
\author[0000-0002-8935-9882]{Akash Anumarlapudi}
\affiliation{Department of Physics and Astronomy, University of North Carolina, Chapel Hill, NC 27599, USA}
\email{akasha@unc.edu}
\author[0000-0003-0638-3340]{Anne M. Archibald}
\affiliation{Newcastle University, NE1 7RU, UK}
\email{anne.archibald@nanograv.org}
\author[0009-0008-6187-8753]{Zaven Arzoumanian}
\affiliation{X-Ray Astrophysics Laboratory, NASA Goddard Space Flight Center, Code 662, Greenbelt, MD 20771, USA}
\email{zaven.arzoumanian@nanograv.org}
\author[0000-0002-4972-1525]{Jeremy G. Baier}
\affiliation{Department of Physics, Oregon State University, Corvallis, OR 97331, USA}
\email{jeremy.baier@nanograv.org}
\author[0000-0003-2745-753X]{Paul T. Baker}
\affiliation{Department of Physics and Astronomy, Widener University, One University Place, Chester, PA 19013, USA}
\email{paul.baker@nanograv.org}
\author[0000-0003-0909-5563]{Bence B\'{e}csy}
\affiliation{Institute for Gravitational Wave Astronomy and School of Physics and Astronomy, University of Birmingham, Edgbaston, Birmingham B15 2TT, UK}
\email{bence.becsy@nanograv.org}
\author[0009-0005-8851-7286]{Amit Bhoonah}
\affiliation{PITT PACC, Department of Physics and Astronomy, University of Pittsburgh, 3941 O’Hara St., Pittsburgh, PA 15260, USA}
\email{amit.bhoonah@nanograv.org}
\author[0000-0002-2183-1087]{Laura Blecha}
\affiliation{Physics Department, University of Florida, Gainesville, FL 32611, USA}
\email{laura.blecha@nanograv.org}
\author[0000-0001-6341-7178]{Adam Brazier}
\affiliation{Cornell Center for Astrophysics and Planetary Science and Department of Astronomy, Cornell University, Ithaca, NY 14853, USA}
\affiliation{Cornell Center for Advanced Computing, Cornell University, Ithaca, NY 14853, USA}
\email{adam.brazier@nanograv.org}
% ...
\shorttitle{Piecewise Power-Law Reconstruction of the Gravitational-Wave Background}
\shortauthors{The NANOGrav Collaboration}

\correspondingauthor{Amit Bhoonah}
{
\let\email\origemail
\email{amit.bhoonah@nanograv.org}
}

\correspondingauthor{Kai Schmitz}
{
\let\email\origemail
\email{kai.schmitz@nanograv.org}}

\begin{abstract}
The NANOGrav 15-year (NG15) data set provides evidence for a gravitational-wave background (GWB) signal at nanohertz frequencies, which is expected to originate either from a cosmic population of inspiraling supermassive black-hole binaries or new particle physics in the early Universe. A firm identification of the source of the NG15 signal requires an accurate reconstruction of its frequency spectrum. In this paper,  we provide such a spectral characterization of the NG15 signal based on a piecewise power-law (PPL) ansatz that strikes a balance between existing alternatives in the literature. Our PPL reconstruction is more flexible than the standard constant-power-law model, which describes the GWB spectrum in terms of only two parameters: an amplitude $A$ and a spectral index $\gamma$. Concurrently, it better approximates physically realistic GWB spectra\,---\,especially those of cosmological origin\,---\,than the free spectral model, since the latter allows for arbitrary variations in the GWB amplitude from one frequency bin to the next. Our PPL reconstruction of the NG15 signal relies on individual PPL models with a fixed number of internal nodes (i.e., constant power law, broken power law, doubly broken power law, etc.) that are ultimately combined in a Bayesian model average. The data products resulting from our analysis provide the basis for fast refits of spectral GWB models.
\end{abstract}



\section{Introduction} 
\label{sec:introduction}

Pulsar timing arrays (PTAs) search for a stochastic gravitational-wave background (GWB) at nHz frequencies by monitoring the times of arrival (TOAs) of pulses from galactic millisecond pulsars over time spans of years and decades~\citep{Taylor:2021yjx}. The hallmark signature of a GWB signal in PTA data consists in a characteristic cross-correlation between the timing residuals of pairs of pulsars, where the timing residuals of each individual pulsar are obtained by comparing its observed TOAs with the predictions of a sophisticated, pulsar-specific timing model. The expected cross-correlation is described by the Hellings--Downs (HD) curve~\citep{Hellings:1983fr}, an overlap reduction function that only depends on the angular separation of pulsars in the sky and which is dominated by its quadrupole contribution when decomposed into Legendre polynomials. The latest generation of PTA data sets\,---\,specifically, the latest data sets from CPTA~\citep{Xu:2023wog}, EPTA and InPTA~\citep{EPTA:2023fyk}, MPTA~\citep{Miles:2024seg}, NANOGrav~\citep{NANOGrav:2023gor}, and PPTA~\citep{Reardon:2023gzh}\,---\,contain an excess timing delay signal that features, at varying levels of statistical significance, the expected HD correlation. It is thus reasonable to assume that PTAs are currently on the brink of a $5\,\sigma$ detection of a GWB.

In this paper, we will focus on the NANOGrav 15-year (NG15) data set~\citep{NANOGrav:2023hde}, which contains evidence for a HD-correlated signal with a statistical significance of $3\cdots4\,\sigma$, and assume that the signal seen in this data set does indeed correspond to a genuine GWB. A crucial question then is that of the origin of the signal. The most plausible source of the signal would arguably be a cosmic population of supermassive black-hole binaries (SMBHBs) at the centers of galaxies~\citep{NANOGrav:2023hfp,DOrazio:2023rvl,NANOGrav:2024nmo,Matt:2025bao}. While no individual SMBHB system close to merger has been detected thus far, it is known that most galaxies host a supermassive black hole at their core as well as that galaxies merge during structure formation. Galaxy mergers across cosmic history should thus give rise to an abundance of SMBHBs and hence a viable source of a GWB signal in the PTA frequency band. Alternatively, the signal may originate from exotic sources in the early Universe necessarily involving new physics beyond the Standard Model of particle physics, e.g., cosmic inflation, scalar-induced gravitational waves (SIGWs), cosmic defects, or a cosmological first-order phase transition~\citep{Caprini:2018mtu,NANOGrav:2023hvm,EPTA:2023xxk,Ellis:2023oxs}. This cosmological interpretation is more speculative than the astrophysical interpretation in terms of SMBHBs. Meanwhile, the detection of a gravitational-wave (GW) echo of the Big Bang would be particularly exciting, as it would open an observational window to the history of our Universe at very early times and hence particle physics at very high energies.


\begin{figure*}
\begin{center}
\includegraphics[width=0.49\textwidth]{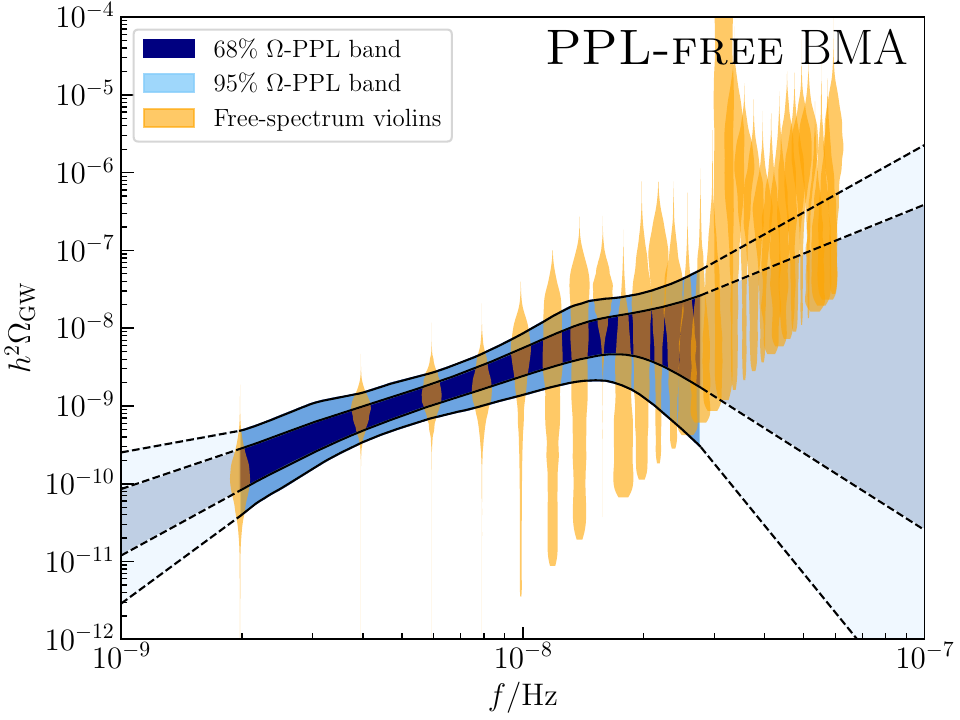}\quad
\includegraphics[width=0.49\textwidth]{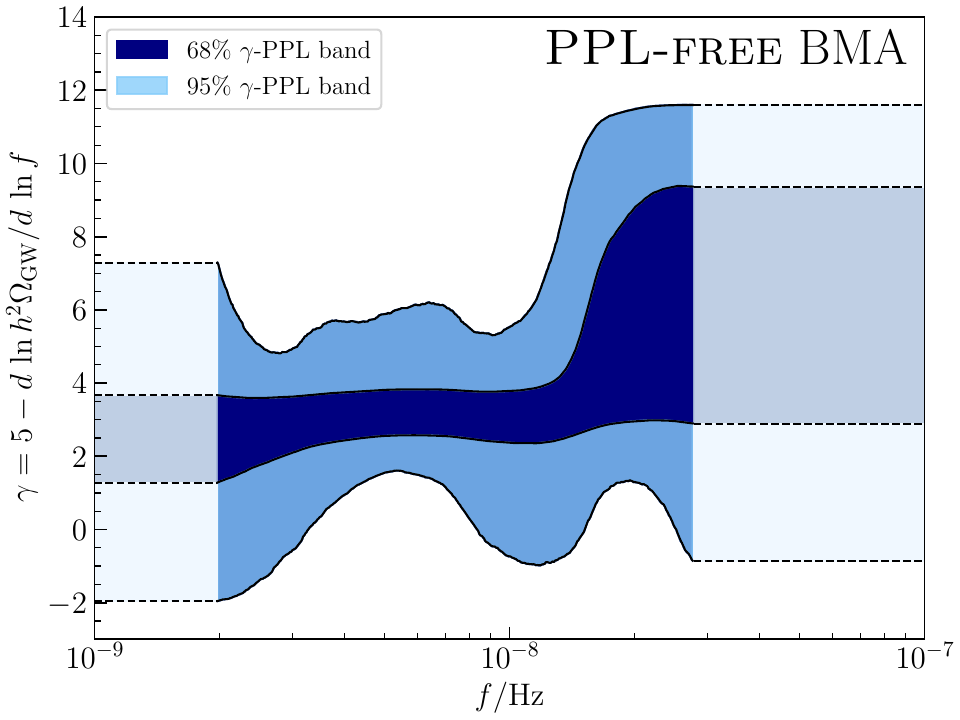}
\end{center}
\caption{\textit{Left panel:} PPL reconstruction of the energy density power spectrum $h^2\Omega_{\rm GW}$ for the GWB signal in the NG15 data, based on the BMA of six PPL models with a fixed number of freely moving internal nodes (\textsc{PPL1-free},\,...\,\textsc{PPL6-free}). \textit{Right panel:} Same, but for the frequency-dependent spectral index of the energy density power spectrum, $\gamma = 5- d \ln h^2\Omega_{\rm GW}/ d\ln f$. Both reconstructions cover the frequency range from $F_1 = \sfrac{1}{T_{\rm obs}}$ to $F_{14} = \sfrac{14}{T_{\rm obs}}$, where $T_{\rm obs} = 16.03\,\textrm{yr}$ (see text for details). In the left panel, we also show the NG15 violins in orange, which represent the amplitude posterior densities for the free spectral model. The lower end of the violins at higher frequencies follows from the prior density chosen in~\cite{NANOGrav:2023gor}.}
\label{fig:omegagammabands}
\end{figure*}


Making progress towards the firm identification of the source of the NG15 signal requires a multi-pronged approach. A PTA observable of particular interest in this context is the GWB angular power spectrum~\citep{NANOGrav:2023tcn,Konstandin:2025ifn}: the GWB signal from SMBHBs is expected to feature much more pronounced anisotropies across the sky than a smooth GWB signal from the early Universe~\citep{Gardiner:2023zzr}. Similarly, searches for continuous-wave (CW) signals from individual SMBHB systems could, in case of a positive signal, confirm that SMBHBs exist and succeed in acting as sources of nHz GWs~\citep{NANOGrav:2023pdq,Agarwal:2025cag}. However, neither GWB anisotropies nor CW signals have been detected thus far, which has two important implications. First, the cosmological interpretation of the NG15 signal remains a viable and relevant option for now. Second, current attempts at model discrimination must focus on yet another PTA observable: the GWB frequency spectrum.  

Spectral GWB models are often formulated in terms of unique model parameters that do not exist in other models, e.g., the cosmic-string tension $G\mu$ in models of GWs from a cosmic-string network, the phase-transition temperature $T_*$ in models of GWs from a cosmological phase transition, etc. To unify the discussion of these different GWB models and facilitate a global model comparison, it is therefore helpful to develop a more general language based on physics-agnostic reference models. The most prominent reference model in this regard is the constant-power-law (CPL) model, which describes the GW energy density power spectrum $h^2\Omega_{\rm GW}$ by a simple power law parametrized in terms of two parameters, an amplitude $A$ and a spectral index $\gamma$ (see below for more details). Furthermore, the CPL model can be easily generalized to a running-power-law (RPL) model by allowing for a logarithmic running of the spectral index, $\gamma \rightarrow \gamma_{\rm run}(f)$~\citep{Agazie:2024kdi}. As recently discussed in \cite{Esmyol:2025ket}, it is straightforward to map beyond-the-Standard-Model (BSM) predictions for the GWB spectrum onto the CPL and RPL models and then use these maps for refitting analyses.

A second prominent physics-agnostic reference model is the free spectral model. While the CPL model can be regarded as a minimal model that fixes the GWB amplitude in each frequency bin in terms of just two parameters, the free spectral model is maximally flexible. Indeed, by definition, the free spectral model treats the GWB amplitude in each frequency bin as a free parameter that can be chosen independently of all other GWB amplitudes. The posterior distributions for these amplitude parameters then yield the well-known NG15 violins that are often used as a graphical illustration of the NG15 signal~\citep{NANOGrav:2023gor}. 

Many physically realistic GWB models are expected to be quite different from these simple reference models. The GWB spectrum from SMBHBs, e.g., can be modeled as a CPL only to first approximation. More realistic SMBHB models predict a spectral turnover at low frequencies due to environmental interactions~\citep{Kocsis:2010xa} and a spectral break at high frequencies caused by the discrete nature of the SMBHB population~\citep{Sesana:2008mz,becsy2022exploring,Agazie:2024jbf,lamb2025finite}. Another example are GWs from cosmic inflation~\citep{Guzzetti:2016mkm}, which do not result in an exact CPL spectrum, even if the primordial tensor power spectrum produced during inflation should have a perfect CPL shape (which by itself is a strong assumption). In this case, the primordial tensor power spectrum will be modulated by the transfer function that relates the primordial spectrum to the present-day GWB spectrum and which is sensitive to changes in the effective number of relativistic degrees of freedom across cosmic history. Together, the primordial tensor spectrum and the transfer function then result in a GWB spectrum that deviates from a CPL~\citep{Watanabe:2006qe,Saikawa:2018rcs}. 

Meanwhile, the free spectral model represents a maximally conservative reconstruction of the GWB signal. Irrespective of physical expectations, it accounts for any conceivable shape of the GWB spectrum, no matter how strongly the GWB amplitude may vary up and down from one frequency bin to the next. It is therefore legitimate to use the free spectral model as a basis for conservative refitting analyses, which is the underlying idea of the \texttt{ceffyl} refitting framework~\citep{Lamb:2023jls}. At the same time, it is clear that the free spectral model encapsulates a multitude of spectral shapes that are unlikely to be realized in any physically realistic model. 

This observation motivates us to seek alternative reference models. We are interested in models that (i) better reflect our physical expectation that the GWB spectrum should be described by a smooth and well-behaved function of frequency, in particular if it originates from a cosmological source, and that (ii) are more flexible than a rigid CPL. The RPL model discussed in \cite{Agazie:2024kdi} can be regarded as a first step in this direction. In the present paper, we continue our construction of physics-agnostic reference models and investigate yet another promising approach: the reconstruction of the GWB signal in terms of a piecewise power law (PPL), i.e., a set of linear-spline interpolations between movable nodes in the $\log_{10}f$--$\log_{10}h^2\Omega_{\rm GW}$ plane.

Our analysis is primarily inspired by literature on the cosmic microwave background (CMB) that is concerned with the reconstruction of the primordial scalar power spectrum~\citep{Vazquez:2012ux,Aslanyan:2014mqa,CORE:2016ymi,Handley:2019fll}. Our reconstruction of the GWB spectrum based on the NG15 data can notably be regarded as the equivalent of the Bayesian reconstruction of the primordial scalar power spectrum by the PLANCK collaboration based on the PLANCK 2015~\citep{Planck:2015sxf} and PLANCK 2018 data~\citep{Planck:2018jri}. Apart from this, the PPL ansatz (sometimes also referred to as the ``nodal'' or ``flexknot'' reconstruction) is routinely used in a variety of analyses in cosmology, ranging from reconstructions of the equation of state of dark energy~\citep{AlbertoVazquez:2012ofj,Hee:2015eba,Ormondroyd:2025iaf} over reconstructions of the cosmic reionization history~\citep{Millea:2018bko,Heimersheim:2021meu} to reconstructions of galaxy cluster profiles~\citep{Olamaie:2017vtt} and the 21 cm signal~\citep{Heimersheim:2023zyi,Shen:2024abc}. In the context of the SIGW interpretation of the NG15 signal, the PPL ansatz was recently also used by \cite{Ghaleb:2025xqn}, in order to constraint the primordial scalar power spectrum in the NANOGrav frequency band.

A decisive advantage of our PPL ansatz is that it properly accounts for the level of complexity supported by the NG15 data. As explained in more detail below, our method relies on a Bayesian model average (BMA) \citep{vanHaasteren:2024lcj} that weights PPL models with different numbers of frequency nodes according to their relative Bayesian evidence. This approach guarantees that our reconstruction of the GWB signal ends up being sufficiently complex, but not more complex than necessary. In addition, we will eventually allow all internal frequency nodes to freely move, such that our final PPL reconstruction of the GWB signal is sensitive to localized features in the GWB spectrum. Conversely, in the absence of such features, our ansatz allows us to constrain local deviations from a smooth CPL-like spectrum; in practice, this is what we will find when applying our methodology to the NG15 data. In summary, the two characteristic properties of our approach\,---\,a data-driven level of complexity and sensitivity to local features (i.e., local constraining power)\,---\,distinguish our ansatz in this work from other physics-agnostic reference models in the literature, such as the CPL or RPL model and the free spectral model. 

The rest of the paper is organized as follows. In Sec.~\ref{sec:analysis}, we will introduce our framework for the reconstruction of the GWB, specifically, our set of PPL models. In particular, we will distinguish between \textsc{binned} and \textsc{free} versions of our models, in which the locations of the internal node frequencies are either kept fixed or allowed to vary. In Sec.~\ref{sec:results}, we will subsequently describe our Bayesian fit to the NG15 data and discuss in turn our main results: first our results for the individual PPL models and then our results for the BMA over these models. Finally, in Sec.~\ref{sec:application}, we will give a brief outlook on how the output of our analysis could be used for refitting analyses in future work, before we conclude in Sec.~\ref{sec:conclusions}. The busy reader may directly skip to Fig.~\ref{fig:omegagammabands}, which shows the main result of our analysis: our final PPL reconstruction of the GWB signal in the NG15 data, alongside the standard NG15 violins. 


\section{Framework and models} 
\label{sec:analysis}


\subsection{GW power spectrum and spectral index}


A GWB induces extra timing delays in the TOAs of the observed pulsars whose covariance can be expressed in terms of the timing-residual cross-power spectrum. For a stochastic, Gaussian, stationary, isotropic, and unpolarized signal in standard Einstein gravity, one finds 
\begin{equation}
\label{eq:SabGW}
S_{ab}^{\rm GWB}\left(f\right) = \Gamma_{ab}\, \frac{S_h\left(f\right)}{6\pi^2f^2} \,.
\end{equation}
Here, $\Gamma_{ab}$ is the cross-correlation coefficient for the pair of pulsars $a$ and $b$, which follows from evaluating the HD overlap reduction function at the angle $\xi_{ab}$, i.e., the angular separation of $a$ and $b$ in the sky,
\begin{equation}
\Gamma_{ab} = \left(1+\delta_{ab}\right) \left[\frac{3}{2}\,x_{ab}\,\ln x_{ab} - \frac{x_{ab}}{4} + \frac{1}{2}\right] \,,
\end{equation}
where $x_{ab}$ is shorthand for $x_{ab} = \sfrac{1}{2}\,(1-\cos\xi_{ab})$. The spectral composition of the signal is accounted for by the GW strain power spectrum $S_h$,\footnote{In our convention, $S_h$ is normalized such that the variance of the tensor metric perturbations $h_{ij}$, summed over the two GW polarization states, can be written as $\langle h_{ij}h^{ij}\rangle = 4\int_0^\infty df\,S_h(f)$.}  which can be related to the characteristic GW strain amplitude 
\begin{equation}
h_c\left(f\right) = \sqrt{2\,f\,S_h\left(f\right)} \,,
\end{equation}
or, alternatively, to the GW energy density power spectrum $h^2\Omega_{\rm GW}$, which measures the fractional GW energy density per logarithmic frequency interval in units of the critical energy density of the present Universe, $\rho_{\rm crit}$,
\begin{equation}
\label{eq:OmegaGW}
h^2\Omega_{\rm GW}\left(f\right) = \frac{1}{\rho_{\rm crit}} \frac{d\rho_{\rm GW}(f)}{d\,\ln f} = \frac{4\pi^2 f^3S_h\left(f\right)}{3\,(H_0/h)^2} \,.
\end{equation}
Here, $H_0$ is the Hubble constant and $h$ its dimensionless version, $H_0 = 100\,h\,\textrm{km}/\textrm{s}/\textrm{Mpc}$. Consequently, $H_0/h$ is just a constant factor that has the units of a frequency and $h^2\Omega_{\rm GW}$ is independent of the observed value of $H_0$. 

Our goal in this paper is to work out a spectral characterization of the GWB signal in the NG15 data, which can be done equally well in terms of $S_h$, $h_c$, and $h^2\Omega_{\rm GW}$. For definiteness, we will present all of our results in the following in terms of $h^2\Omega_{\rm GW}$, which represents the standard choice in large parts of the cosmology and early-Universe literature. In addition, we will be interested in a reconstruction of the frequency-dependent index of the GW energy density power spectrum $h^2\Omega_{\rm GW}$,
\begin{equation}
n\left(f\right) = \frac{d\,\ln h^2\Omega_{\rm GW}\left(f\right)}{d\,\ln f} \,,
\end{equation}
which is closely related to the spectral index of the timing-residual cross power spectrum in Eq.~\eqref{eq:SabGW},
\begin{equation}
\gamma\left(f\right) = -\frac{d\,\ln S_{ab}^{\rm GW}\left(f\right)}{d\,\ln f} = 5 - n\left(f\right) \,,
\end{equation}
where the negative sign in the first identity is a standard convention. For definiteness and in order to facilitate the comparison with other parts of the PTA literature, we will report our results for the spectral index in terms of $\gamma$ rather than $n$ in the remainder of this paper.


\subsection{PPL models}

Let us now define the PPL models that our spectral reconstruction of the GWB is going to be based on. The simplest PPL model is equivalent to the CPL model,
\begin{equation}
\label{eq:CPL}
h^2\Omega_{\rm GW}\left(f\right) \:\:\overset{\rm CPL}{=}\:\: h^2\Omega_{\rm GW}\left(f_0\right) \left(\frac{f}{f_0}\right)^{5-\gamma_1} \,,
\end{equation}
for some reference frequency $f_0$. This model has two free parameters: the amplitude of $h^2\Omega_{\rm GW}$ at the reference frequency, $h^2\Omega_{\rm GW}\left(f_0\right)$, and the spectral index $\gamma_1$, which takes the same constant value at all frequencies. In our Bayesian fit analyses below, we will account for the presence of a GWB signal in the frequency range from $f_{\rm min} = \sfrac{1}{T_{\rm obs}}$ to  $f_{\rm max} = \sfrac{14}{T_{\rm obs}}$, where $T_{\rm obs}$ denotes the total observing time span, $T_{\rm obs} = 16.03\,\textrm{yr}$. This frequency range corresponds to the first 14 frequency bins in the discrete Fourier series that we shall employ to model the NG15 data. In the simplest PPL model introduced in Eq.~\eqref{eq:CPL}, but also in all other PPL models, we will fix the reference frequency $f_0$ that determines the overall amplitude factor $h^2\Omega_{\rm GW}\left(f_0\right)$ at $f_0 = f_{\rm min}$. 

The next-simplest PPL model describes the GW power spectrum in terms of a broken power law (BPL),
\begin{align}
\label{eq:BPL}
& h^2\Omega_{\rm GW}\left(f\right) \:\:\overset{\rm BPL}{=}\:\: h^2\Omega_{\rm GW}\left(f_0\right)  \\
\nonumber
& \times \begin{cases} \left(\frac{f}{f_0}\right)^{5-\gamma_1} &,\quad f_{\rm min} \leq f \leq f_1 \\ \left(\frac{f_1}{f_0}\right)^{5-\gamma_1}\left(\frac{f}{f_1}\right)^{5-\gamma_2} & ,\quad  f_1 \leq f \leq f_{\rm max} \end{cases} \,,
\end{align}
which now features one amplitude, $h^2\Omega_{\rm GW}\left(f_0\right)$, and two spectral indices, $\gamma_1$ and $\gamma_2$, which describe the power-law behavior of the GW spectrum at $f < f_1$ and $f > f_1$, respectively. The frequency $f_1$ marks the position of the spectral break in the BPL model, i.e., the internal node in this PPL model.

For all PPL models with at least two spectral indices, we shall distinguish between two model variants in the following: a \textsc{binned} and a \textsc{free} version. In the \textsc{binned} models, we do not treat the location of the node frequencies as free parameters. Instead, we keep them fixed at equidistant separations on a logarithmic frequency axis in order to set up equally sized frequency bins spanning the range from $f_{\rm min}$ to $f_{\rm max}$. In the case of the BPL model introduced above, this results in what we shall refer to as the \textsc{PPL2-binned} model, in which $f_0 = f_{\rm min} = \sfrac{1}{T_{\rm obs}}$, $f_1 = \sqrt{f_0\,f_2} = \sfrac{\sqrt{14}}{T_{\rm obs}}$, and $f_2 = f_{\rm max} = \sfrac{14}{T_{\rm obs}}$. In this model, we thus work with frequency bins for the spectral index: the frequency-dependent index $\gamma$ is set to $\gamma_1$ in the bin from $f_0$ to $f_1$ and to $\gamma_2$ in the bin from $f_1$ to $f_2$. By contrast, in the \textsc{free} versions of our PPL models, we allow the position of the internal node frequencies to vary freely. For the BPL model introduced above, this results in what we shall refer to as the \textsc{PPL2-free} model, in which only $f_0$ and $f_2$ are kept fixed, but $f_1$ is allowed to vary between $f_0$ and $f_2$.  In summary, this means that the \textsc{PPL2-binned} model features three model parameters, $\left\{h^2\Omega_{\rm GW}\left(f_0\right),\gamma_1,\gamma_2\right\}$, while the \textsc{PPL2-free} model features four model parameters, $\left\{h^2\Omega_{\rm GW}\left(f_0\right),\gamma_1,\gamma_2,f_1\right\}$. Note that a similar distinction is not possible for the CPL model, i.e., the \textsc{PPL1} model, which does not feature any internal node. Therefore, when we use model names such as \textsc{PPL1-free} (see the caption of Fig.~\ref{fig:omegagammabands}) and \textsc{PPL1-binned} (see the caption of Fig.~\ref{fig:og-binned-bma}), it should be kept in mind that all three names (\textsc{PPL1-free}, \textsc{PPL1-binned}, \textsc{PPL1}) denote the same model. 

Below, our analysis of the \textsc{binned} PPL models will serve as a warm-up for our analysis of the \textsc{free} PPL models. The \textsc{binned} PPL models are slightly easier to handle because of the lower dimensionality of their parameter space. In particular, this will allow us to avoid parameter degeneracies, i.e., flat directions in two-dimensional (2D) marginalized posterior densities. In addition, the \textsc{binned} models are inspired by related approaches in the literature; see, e.g., \cite{Caprini:2019pxz}, which discusses a method for the spectral reconstruction of the GWB signal that will be measured by the Laser Interferometer Space Antenna (LISA). The method described by \cite{Caprini:2019pxz} also relies on a bin-by-bin power-law reconstruction of the GWB signal (i.e., a binned PPL reconstruction) and provides the basis for the dedicated software package \texttt{SGWBinner}. At the same time, it is clear that the \textsc{binned} PPL models are more restricted than the \textsc{free} PPL models. Our most general and flexible reconstruction of the GWB signal will therefore be based on the \textsc{free} PPL models, ensuring that the data itself decide where breaks or features need to be placed on the frequency axis. 

Before we move on to the next PPL model, we would like to comment on an alternative possible parametrization of the \textsc{PPL2-binned} and \textsc{PPL2-free} models. In our analysis, we deliberately construct our parametrization around the spectral indices $\gamma_1$ and $\gamma_2$, as this will allow us to directly reconstruct the GW power spectrum and its frequency-dependent spectral index simultaneously. Alternatively, both models can also be parametrized in terms of the amplitudes at the boundaries of the frequency band, $h^2\Omega_{\rm GW}\left(f_0\right)$ and $h^2\Omega_{\rm GW}\left(f_2\right)$, and the location of the internal node, $\left\{f_1,h^2\Omega_{\rm GW}\left(f_1\right)\right\}$, where $f_1$ is either treated as a free parameter (\textsc{PPL2-free}) or not (\textsc{PPL2-binned}). The relation between these two possible parameterizations is straightforward and directly follows from Eq.~\eqref{eq:BPL}, 
\begin{align}
h^2\Omega_{\rm GW}\left(f_1\right) & = h^2\Omega_{\rm GW}\left(f_0\right)\left(\frac{f_1}{f_0}\right)^{5-\gamma_1} \,, \\
h^2\Omega_{\rm GW}\left(f_2\right) & = h^2\Omega_{\rm GW}\left(f_1\right)\left(\frac{f_2}{f_1}\right)^{5-\gamma_2} \,.
\end{align}
In their reconstruction of the primordial scalar power spectrum~\citep{Planck:2015sxf,Planck:2018jri}, the PLANCK collaboration works with a parametrization of the second type, i.e., in terms of amplitudes and frequencies rather than spectral indices and frequencies. From the perspective of the Bayesian inference, both approaches are equivalent as long as the prior densities of all relevant parameters are chosen accordingly. In our analysis, we will work, e.g., with uniform priors on all spectral indices and log-uniform priors on all amplitudes and frequencies. For appropriately chosen prior ranges, our analysis can thus be mapped onto an analysis that parametrizes the GW power spectrum exclusively in terms of amplitudes and frequencies and that uses log-uniform priors for all these parameters. 

The next PPL model describes the GW power spectrum in terms of a doubly broken power law (DBPL), 
\begin{align}
\label{eq:DBPL}
& h^2\Omega_{\rm GW}\left(f\right) \:\:\overset{\rm DBPL}{=}\:\: h^2\Omega_{\rm GW}\left(f_0\right)\\
\nonumber
&  \times \begin{cases} \left(\frac{f}{f_0}\right)^{5-\gamma_1} & ,\quad f_{\rm min} \leq f \leq f_1 \\ \left(\frac{f_1}{f_0}\right)^{5-\gamma_1}\left(\frac{f}{f_1}\right)^{5-\gamma_2} & ,\quad  f_1 \leq f \leq f_2 \\ \left(\frac{f_1}{f_0}\right)^{5-\gamma_1}\left(\frac{f_2}{f_1}\right)^{5-\gamma_2}\left(\frac{f}{f_2}\right)^{5-\gamma_3} & ,\quad  f_2 \leq f \leq f_{\rm max} \end{cases} \,.
\end{align}
Again, we distinguish between two versions of the model: \textsc{PPL3-binned} and \textsc{PPL3-free}, where the former features four parameters, $\left\{h^2\Omega_{\rm GW}\left(f_0\right),\gamma_1,\gamma_2,\gamma_3\right\}$, and the latter six parameters, $\left\{h^2\Omega_{\rm GW}\left(f_0\right),\gamma_1,\gamma_2,\gamma_3,f_1,f_2\right\}$. In the \textsc{PPL3-binned} model, the two internal node frequencies are fixed at $f_1 = \sfrac{14^{1/3}}{T_{\rm obs}}$ and $f_2 = \sfrac{14^{2/3}}{T_{\rm obs}}$. Similarly as in the case of a BPL, we could switch again to an alternative parametrization purely in terms of node amplitudes $h^2\Omega_{\rm GW}\left(f_i\right)$ and node frequencies $f_i$.

At this point, it is now obvious how to generalize the PPL models introduced thus far to PPL models with an even larger number of internal nodes, i.e., to a triply broken power law, a quadruply broken power law, etc. For a general \textsc{PPL$n$} model, featuring in total $n$ different spectral indices $\gamma_i$ with $i=1,\cdots,n$, we may write
\begin{align}
\label{eq:PPLn}
& h^2\Omega_{\rm GW}\left(f\right) \:\:\overset{\textrm{PPL$n$}}{=}\:\: h^2\Omega_{\rm GW}\left(f_0\right) \\
& \times \left[\prod_{i=1}^{m-1} \left(\frac{f_i}{f_{i-1}}\right)^{5-\gamma_i}\right] \left(\frac{f}{f_{m-1}}\right)^{5-\gamma_m} \,,
\end{align}
when $f$ falls into the interval $[f_{m-1},f_m]$. The \textrm{PPL$n$} model features $n$ such intervals, $m=1,\cdots,n$. In the \textsc{free} version of the model, the internal node frequencies $f_i$ with $i=1,\cdots,n-1$ are treated as free parameters that are only subject to the requirement that they be correctly ordered, $f_{\rm min} = f_0 \leq f_1 \leq f_2 \leq \cdots \leq f_n = f_{\rm max}$. In the \textsc{binned} version of the model, on the other hand, all internal frequencies are fixed at $f_i = \sfrac{14^{i/n}}{T_{\rm obs}}$. The two nodes at the boundary of the frequency range are always kept constant in both models, $f_0 = \sfrac{1}{T_{\rm obs}}$ and $f_n = \sfrac{14}{T_{\rm obs}}$. The \textsc{PPL$n$-binned} model thus features $n+1$ parameters, $\left\{h^2\Omega_{\rm GW}\left(f_0\right),\gamma_1,\cdots,\gamma_n\right\}$, while the \textsc{PPL$n$-free} model features $(n+1)+(n-1)= 2n$ parameters, $\left\{h^2\Omega_{\rm GW}\left(f_0\right),\gamma_1,\cdots,\gamma_n, f_1,\cdots,f_{n-1}\right\}$. 


\begin{table*}
\renewcommand{\arraystretch}{1.2}

\caption{Prior probability density distributions for the free parameters in our fit of the PPL models to the NG15 data.}
\label{tab:priors}

\begin{flushleft}
\begin{tabular}{llll}
\toprule
\textbf{Parameter} & \textbf{Description} & \textbf{Prior} & \textbf{Comments}                          \\ [4pt] \midrule
\multicolumn{4}{c}{\textbf{Pulsar-intrinsic red noise}}                                                 \\
$A_a^{\rm red}$      &amplitude                      & log-uniform $[-20,-11]$ & one parameter per pulsar $a$ \\
$\gamma_a^{\rm red}$ & spectral index            & uniform $[0,7]$         & one parameter per pulsar $a$ \\ [4pt] \midrule
\multicolumn{4}{c}{\textbf{GWB in the \textsc{PPL\boldmath{$n$}-binned} models}} \\ 
$h^2\Omega_{\rm GW}\left(\sfrac{1}{T_{\rm obs}}\right)$ & amplitude         & log-uniform $[-12,-8]$ & one parameter per PTA      \\
$\gamma_i$ with $i=1,\cdots,n$                          & spectral indices  & uniform $[-3,12]$        & $n$ parameters per PTA     \\
\midrule 
\multicolumn{4}{c}{\textbf{GWB in the \textsc{PPL\boldmath{$n$}-free} models}} \\ 
$h^2\Omega_{\rm GW}\left(\sfrac{1}{T_{\rm obs}}\right)$ & amplitude         & log-uniform $[-12,-8]$ & one parameter per PTA      \\
$\gamma_i$ with $i=1,\cdots,n$                          & spectral indices  & uniform $[-3,12]$        & $n$ parameters per PTA     \\
$f_i$ with $i=1,\cdots,n-1$                             & node frequencies  & log-uniform $[\log_{10}\sfrac{1}{T_{\rm obs}},\log_{10}\sfrac{14}{T_{\rm obs}}]^\dagger$ & $n-1$ parameters per PTA   \\
\bottomrule 
\end{tabular}
\end{flushleft}

\hfill\footnotesize{${}^\dagger$\quad After drawing the node frequencies from this prior density, we need to sort them according to their size (see text for more details).}
\end{table*}


\section{Bayesian fit to the NG15 data} 
\label{sec:results}


\subsection{PTA likelihood and prior choices}

We are now ready to fit the \textsc{PPL$n$-binned} and \textsc{PPL$n$-free} models to the NG15 data. This data set spans a total observing time $T_{\rm obs}\simeq 16.03\,\textrm{yr}$ and features pulse TOAs from 68 millisecond pulsars~\citep{NANOGrav:2023hde}. One of these pulsars was observed for less than three years and is not included in our analysis. The Bayesian fit framework employed in this paper is closely related to the methods used in our earlier work; see our search for signals from new physics~\citep{NANOGrav:2023hvm} and our fit of the RPL model to the NG15 data~\citep{Agazie:2024kdi}. In the following, we will therefore only briefly summarize the key ingredients of our analysis framework and provide more details only when certain aspects of our analysis differ from our approach in earlier work. 

The central observables at the heart of our fit analysis are the NG15 timing residuals, which follow from the observed TOAs after subtracting the expected TOAs predicted by the respective timing models. In our Bayesian fit framework, we model these timing residuals in terms of three different contributions: (i) linear offsets from the best-fit values of the timing-model parameters, (ii) frequency-independent white noise, and (iii) frequency-dependent red noise. On top, we model pulse dispersion, again following~\cite{NANOGrav:2023hvm}, as a piecewise constant with the inclusion of DMX parameters~\citep{NANOGrav:2015qfw,Jones:2016fkk}.

Similar to our earlier work, we marginalize over the timing-model uncertainties and keep the white-noise parameters fixed at the maximum-posterior values that were previously determined in single-pulsar analyses~\citep{NANOGrav:2023ctt}. The physically interesting contributions to the timing residuals reside in the red-noise terms, which encompass pulsar-intrinsic red noise and the GWB signal. More specifically, the GWB signal is represented by a common-spectrum red-noise term with HD cross-correlation coefficients $\Gamma_{ab}$. In the limit of trivial cross-correlation coefficients, $\Gamma_{ab} \rightarrow \delta_{ab}$, the GWB signal reduces to a common-spectrum uncorrelated red-noise (CURN) signal. We will come back to the relation between these two types of signals (GWB and CURN) in our discussion on Bayes factors below. 

We model both types of red noise, pulsar-intrinsic red noise and the GWB signal, in terms of discrete Fourier series with harmonic frequencies $F_k = \sfrac{k}{T_{\rm obs}}$.%
\footnote{Here, we choose a notation that clearly distinguishes between the node frequencies in our PPL models, $f_i = \sfrac{14^{i/n}}{T_{\rm obs}}$, and the harmonic frequencies in the Fourier series, $F_k = \sfrac{k}{T_{\rm obs}}$.}
The intrinsic red noise of each pulsar is described by a power law, which amounts to two red-noise parameters per pulsar $a$: an amplitude $A_a^{\rm red}$ and a spectral index $\gamma_a^{\rm red}$. The GWB signal, on the other hand, is modeled by the timing-residual cross-power spectrum in Eq.~\eqref{eq:SabGW}, which, together with the relation in Eq.~\eqref{eq:OmegaGW}, can be written as 
\begin{equation}
\label{eq:SGWB}
S_{ab}^{\rm GWB}\left(f\right) = \Gamma_{ab}\,\frac{(H_0/h)^2}{8\pi^4 f^5}\,h^2\Omega_{\rm GW}\left(f\right) \,.
\end{equation}
Here, $h^2\Omega_{\rm GW}$ needs to be chosen accordingly to the PPL model under consideration. Moreover, the GWB contribution to the discrete Fourier series is only sensitive to the values of $S_{ab}^{\rm GWB}$ at $f = F_k$. We include pulsar-intrinsic red noise at all harmonic frequencies across the whole NANOGrav frequency band from $F_1$ to $F_{30}$, but restrict the GWB signal to the first 14 harmonic frequencies from $F_1$ to $F_{14}$. This is consistent with the observation that the GWB signal in the NG15 data mostly appears at lower frequencies~\citep{NANOGrav:2023gor}.

The power spectrum in Eq.~\eqref{eq:SGWB}, alongside the power spectrum for pulsar-intrinsic red noise, determines the covariance matrix of the coefficients in our discrete Fourier series. We are not interested in the explicit realization of the red noise in the NG15 data set, only its stochastic properties that are described by the parameters (or more precisely, hyperparameters) entering the covariance matrix. We, therefore, marginalize the likelihood for the timing residuals not only over the timing-model uncertainties, but also over the Fourier-series coefficients. This procedure results in the standard marginalized PTA likelihood~\citep{vanHaasteren:2012hj,Lentati:2012xb}, which forms the basis of our numerical data analysis. 

The marginalized PTA likelihood quantifies the probability $\mathcal{L}(D|\bm{\theta})$ of the NG15 data $D$ as a function of the  hyperparameters $\bm{\theta}$, which encompass 134 parameters for pulsar-intrinsic red noise (67 amplitudes $A_a^{\rm red}$ and 67 spectral indices $\gamma_a^{\rm red}$) as well as $n+1$ GWB parameters in the case of the \textsc{PPL$n$-binned} models or $2n$ GWB parameters in the case of the \textsc{PPL$n$-free} models. In our Bayesian fit analysis, we need to combine the likelihood $\mathcal{L}(D|\bm{\theta})$ with prior densities for all model parameters $\bm{\theta}$. Our choices for these prior densities are summarized in Table~\ref{tab:priors}. As discussed before, we work with uniform priors for all spectral indices as well as log-uniform priors for all amplitude and frequency parameters. In particular, our prior ranges are chosen large enough such that we are typically able to capture the entire relevant posterior volume in the high-dimensional parameter space. At the same time, our prior choice for the spectral indices $\gamma_i$ is supposed to reflect that we consider even smaller or larger indices, i.e., a GW spectrum scaling like $h^2\Omega_{\rm GW} \propto f^n$ with $n>8$ or $n< -7$, highly improbable from a physical perspective. The boundaries of this prior range are somewhat arbitrary, which represents a typical issue in Bayesian statistics. On the other hand, our choice for the $\gamma_i$ prior densities is certainly a conservative one, which allows us to explore the bulk of the posterior density, even in the presence of parameter degeneracies in the \textsc{PPL$n$-free} models. Moreover, a second important factor to consider is that our prior choices are universal across all PPL models under investigation, which is essential for a fair model comparison and the determination of meaningful Bayes factors. 

The node frequencies $f_i$ are first drawn from log-uniform prior ranges as indicated in Table~\ref{tab:priors} and then sorted according to their size, $f_1 \leq f_2 \leq \cdots \leq f_{n-1}$, before they are used to evaluate the GW power spectrum $h^2\Omega_{\rm GW}$. At the level of the posterior density, this algorithm effectively amounts to an analysis in which we simultaneously explore several identical copies of the posterior density that are eventually mapped onto one common posterior density. Consider, e.g., the PPL3 model, which features two break frequencies, $f_1$ and $f_2$. The posterior density that we would like to explore lives in the region of the $f_1$--$f_2$ parameter space where $f_ 1\leq f_2$. We can, however, continue this posterior density to the region where $f_1 > f_2$ by exchanging $f_1$ and $f_2$ in the likelihood. From this, we obtain a mirrored copy of the original posterior density in the second half of parameter space that we can combine with the original posterior density to obtain one final posterior density. In short, first drawing the node frequencies from independent prior ranges as in Table~\ref{tab:priors} and then sorting them allows us to avoid a more complicated construction of a multidimensional prior density that accounts for the correct order of the node frequencies from the start. 

In passing, we also stress that our treatment of the node frequencies leaves the marginal likelihoods of the PPL models unaffected, which means that it does not distort the Bayes factors that we will need later on in our analysis. Indeed, for PPL$n$ and $n \geq 3$, our analysis operates on $(n-1)!$ copies of the relevant parameter-space volume. But at the same time, we renormalize the prior density for the internal node frequencies $f_i$, such that it integrates to unity, not on the initial parameter space where all node frequencies are sorted, but on the whole enlarged parameter space. This introduces a factor of $\sfrac{1}{(n-1)!}$ at the level of the prior density, which cancels the $(n-1)!$ factor in the parameter-space volume. In practice, this cancellation takes places automatically. We simply have to make sure that the prior density for the internal node frequencies $f_i$ is normalized to unity on the full $[\log_{10}\sfrac{1}{T_{\rm obs}},\log_{10}\sfrac{14}{T_{\rm obs}}]^{n-1}$ hypercube, which is exactly what we do in our analysis (see Tab.~\ref{tab:priors}).  

With the PTA likelihood and prior choices in Table~\ref{tab:priors} in place, we are now ready to sample from the posterior densities of our PPL models. We do so using standard Markov chain Monte Carlo (MCMC) techniques~\citep{justin_ellis_2017_1037579} provided by the software package \texttt{PTArcade}~\citep{Mitridate:2023oar}, a wrapper for \texttt{ENTERPRISE}~\citep{2019ascl.soft12015E} and \texttt{ENTERPRISE\_EXTENSIONS}~\citep{enterprise}. Our main results presented below are based on MCMC runs in \texttt{PTArcade}'s \textit{enterprise} mode, i.e., Bayesian fits based on the PTA likelihood for the timing residuals. However, for testing and validation purposes, we also make use of \texttt{PTArcade}'s \textit{ceffyl} mode that allows one to perform Bayesian fits based on a factorized likelihood that is induced by the posterior densities of the free spectral model (i.e., by the NG15 violins)~\citep{Lamb:2023jls}. 


\begin{figure*}
\begin{center}
\includegraphics[width=0.23\textwidth]{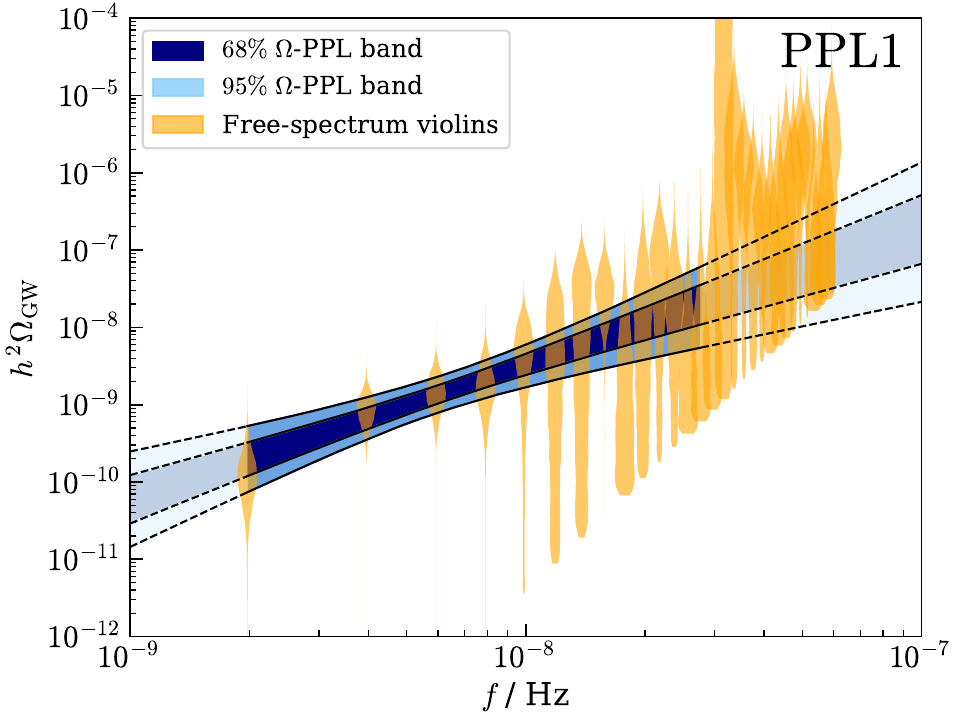}\quad
\includegraphics[width=0.23\textwidth]{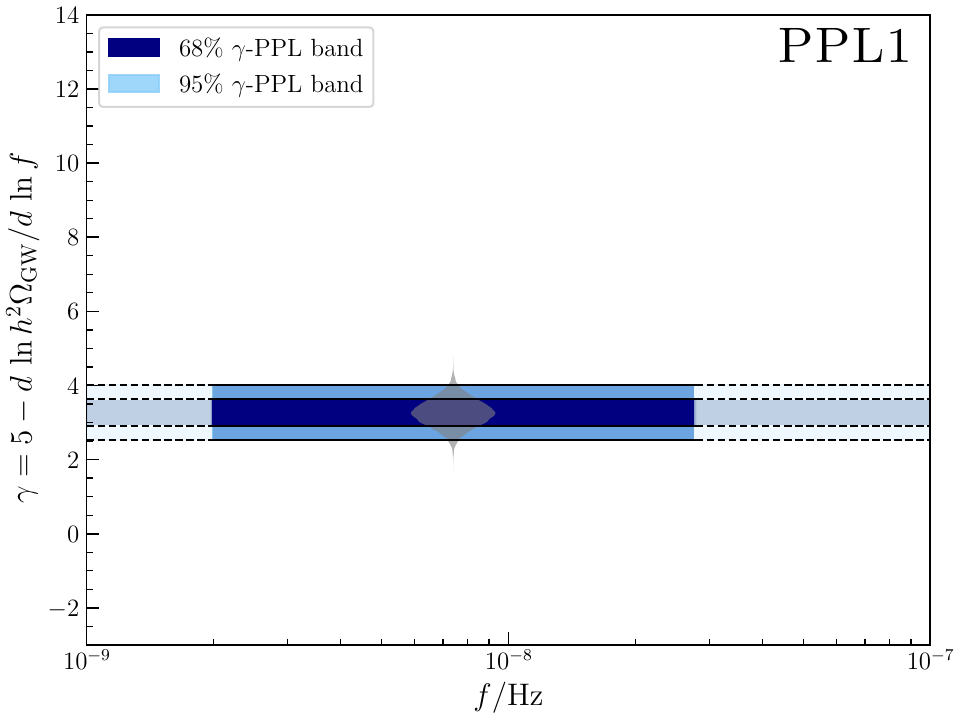} \quad
\includegraphics[width=0.23\textwidth]{fig/Omega_PPL_1_Band.pdf}\quad
\includegraphics[width=0.23\textwidth]{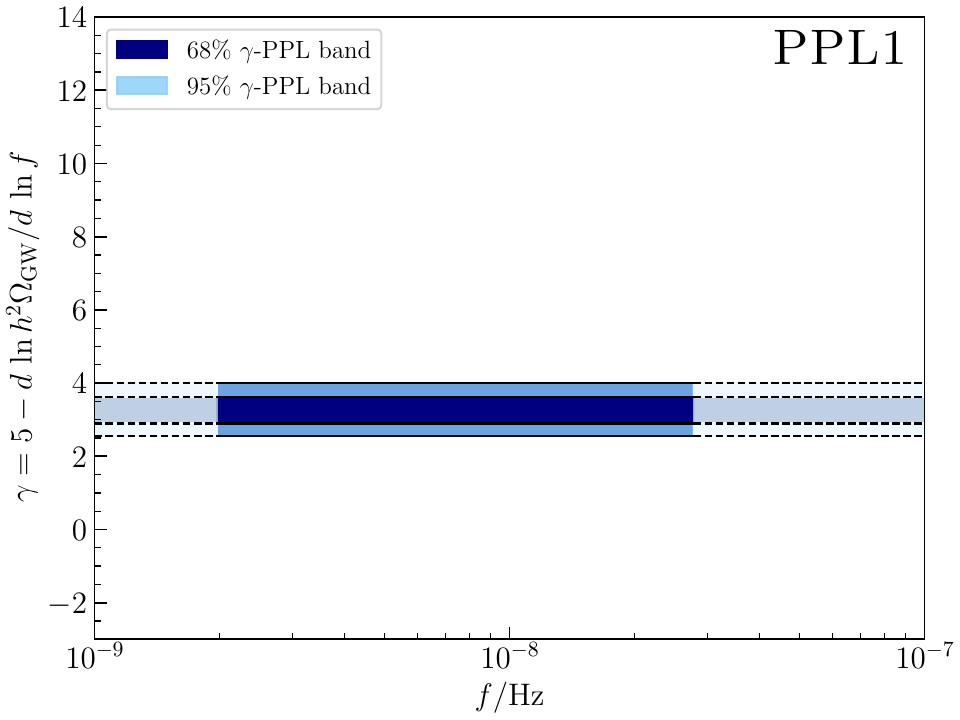}

\includegraphics[width=0.23\textwidth]{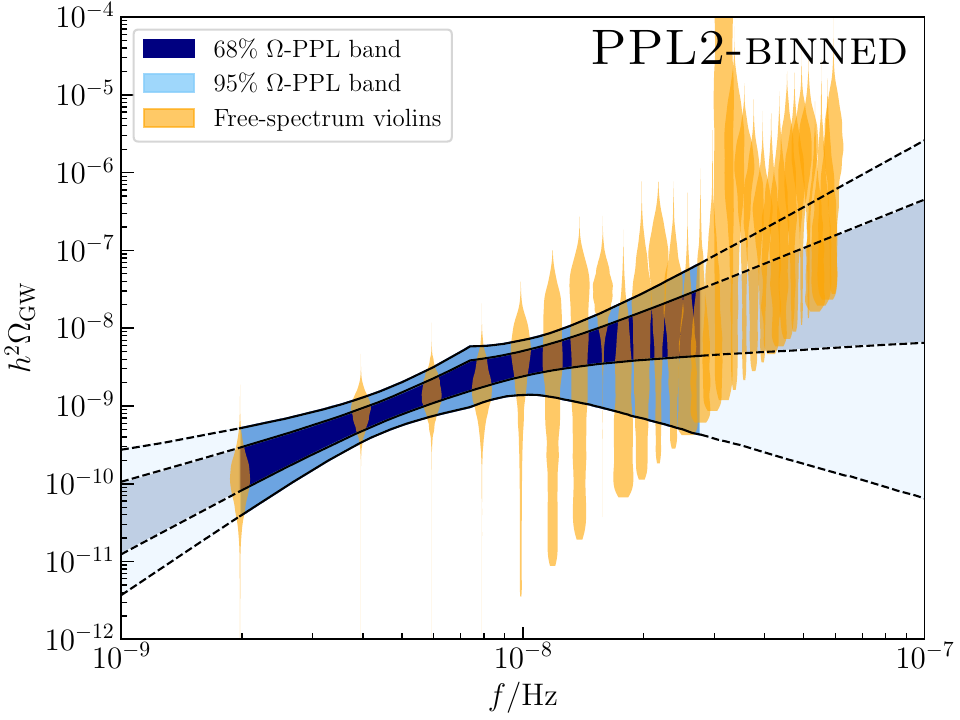}\quad
\includegraphics[width=0.23\textwidth]{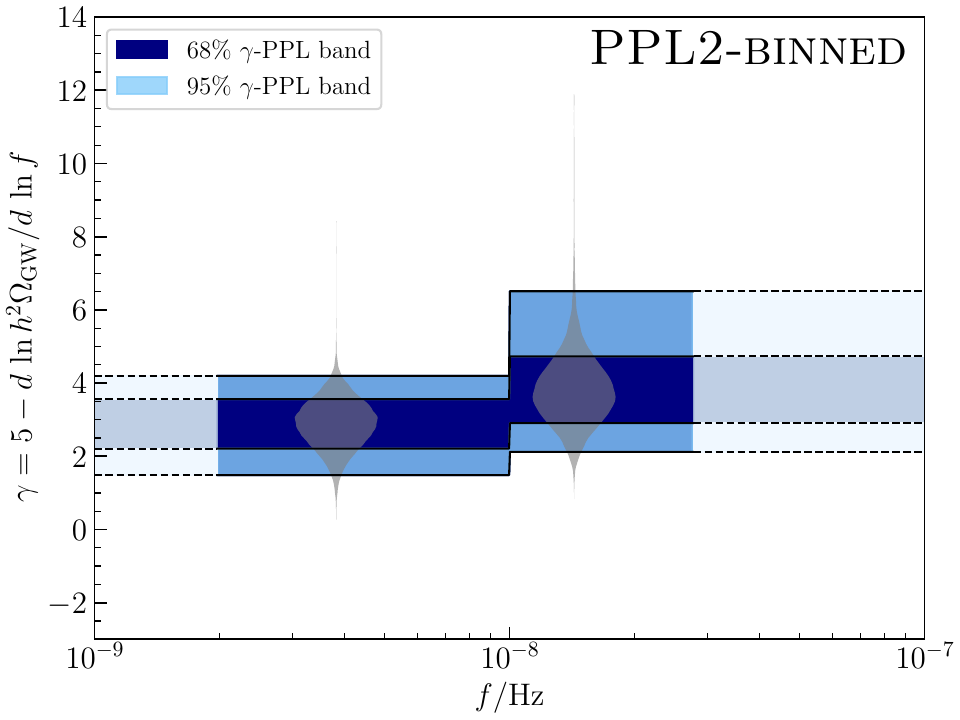} \quad
\includegraphics[width=0.23\textwidth]{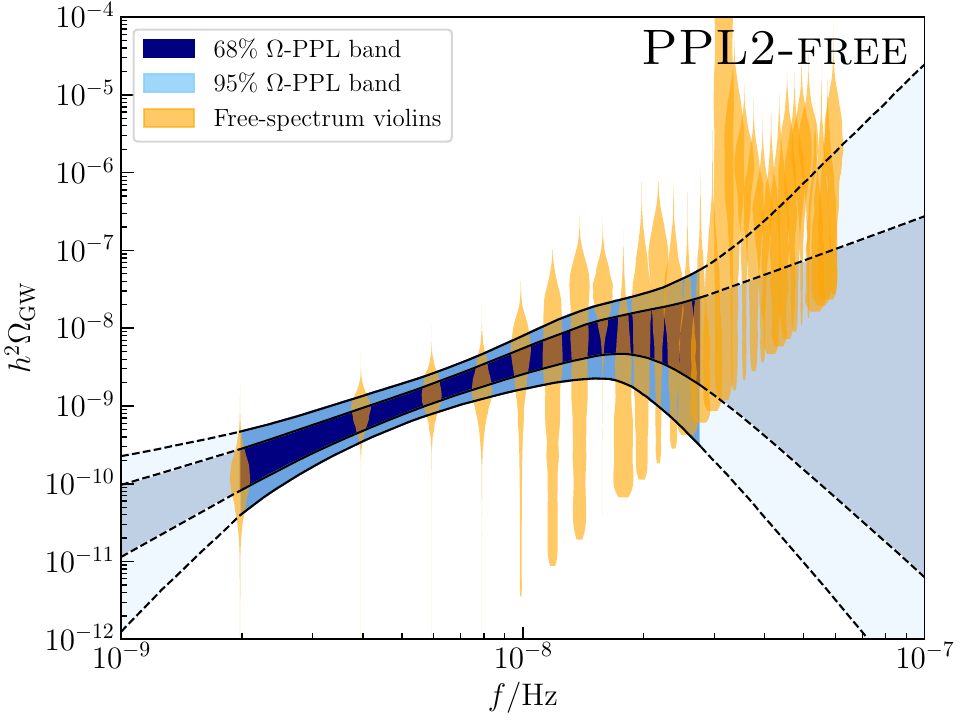}\quad
\includegraphics[width=0.23\textwidth]{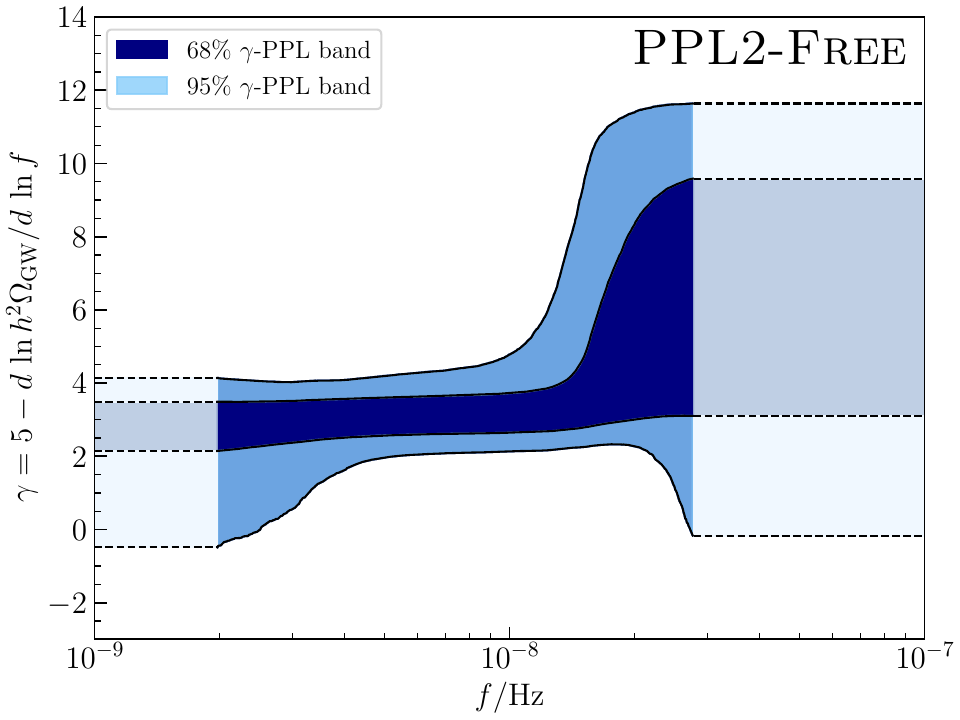}

\includegraphics[width=0.23\textwidth]{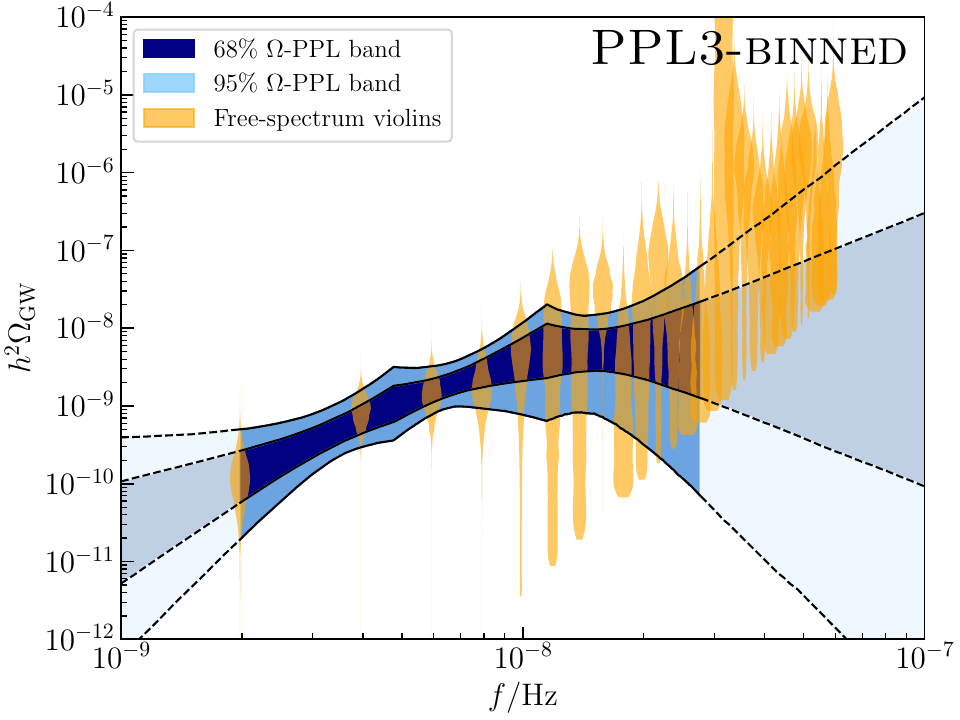}\quad
\includegraphics[width=0.23\textwidth]{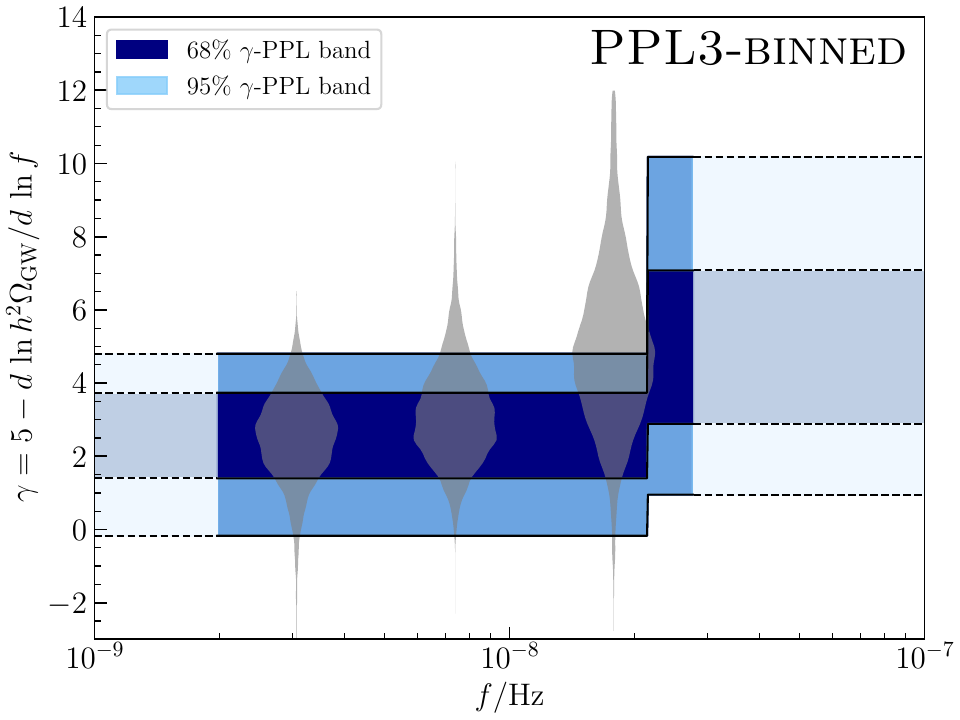} \quad
\includegraphics[width=0.23\textwidth]{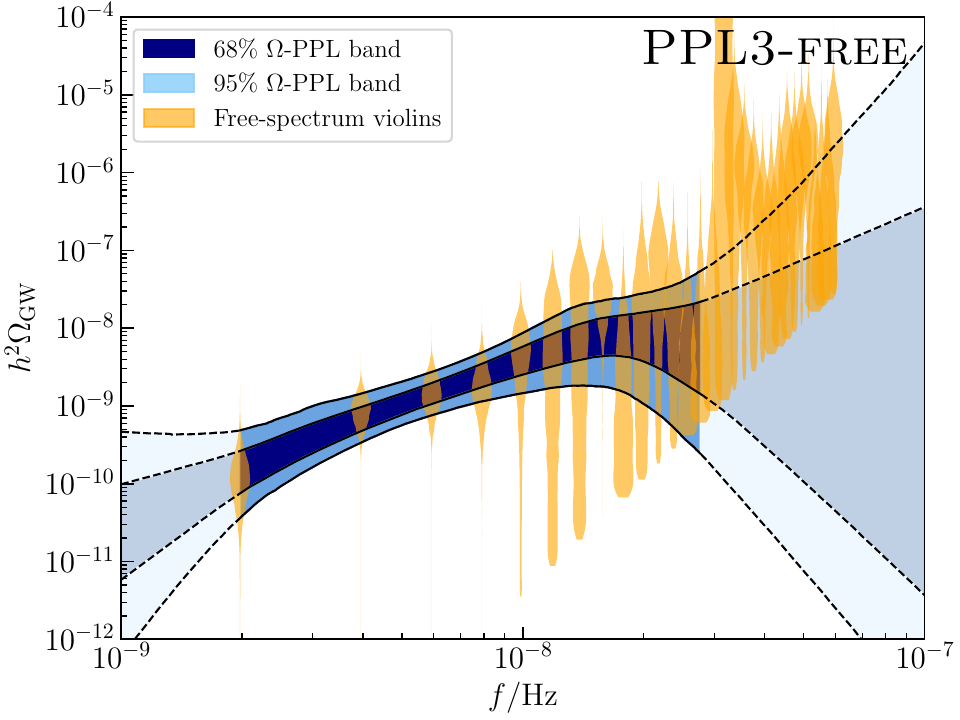}\quad
\includegraphics[width=0.23\textwidth]{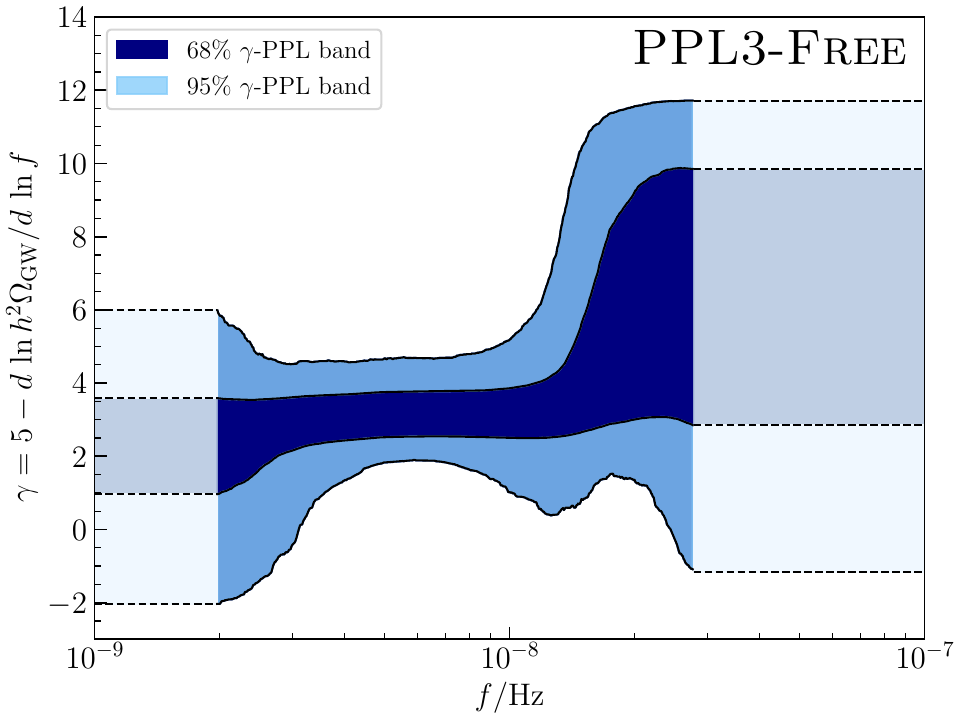}

\includegraphics[width=0.23\textwidth]{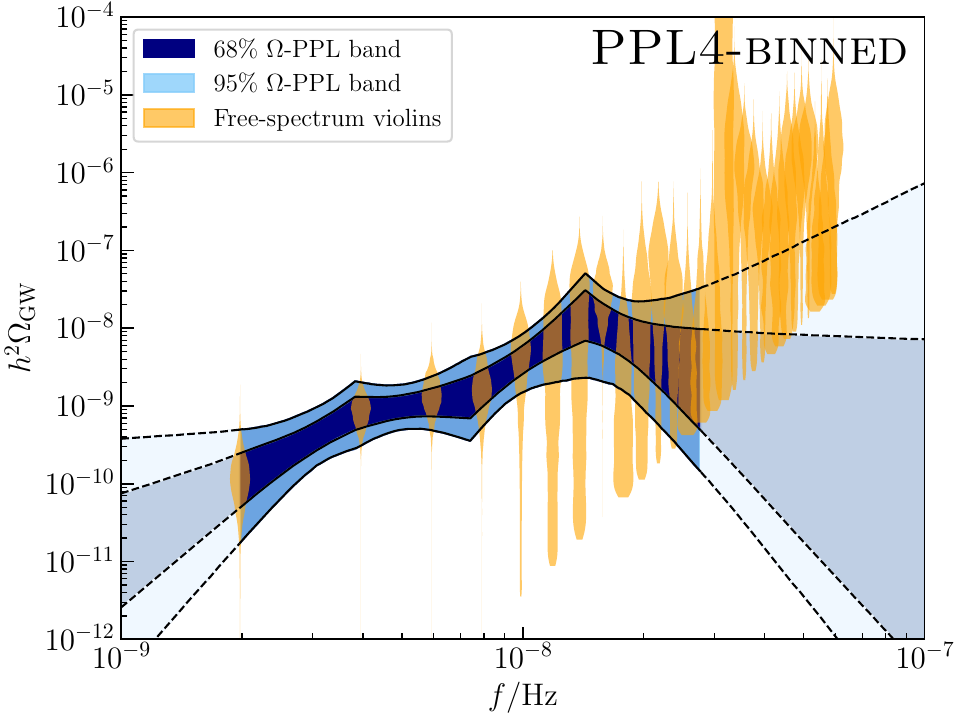}\quad
\includegraphics[width=0.23\textwidth]{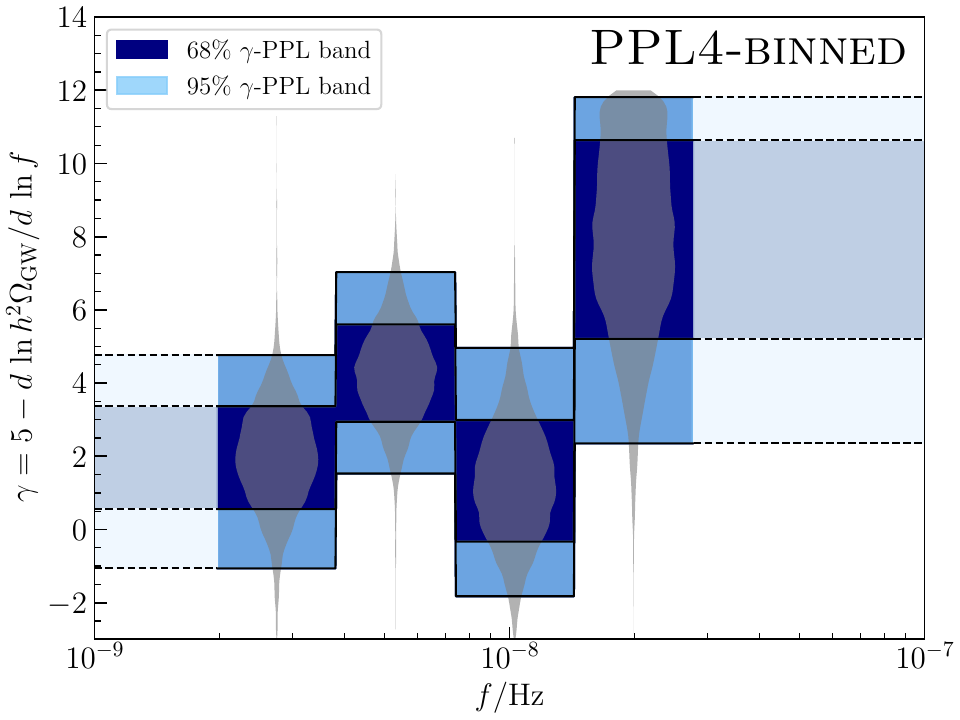} \quad
\includegraphics[width=0.23\textwidth]{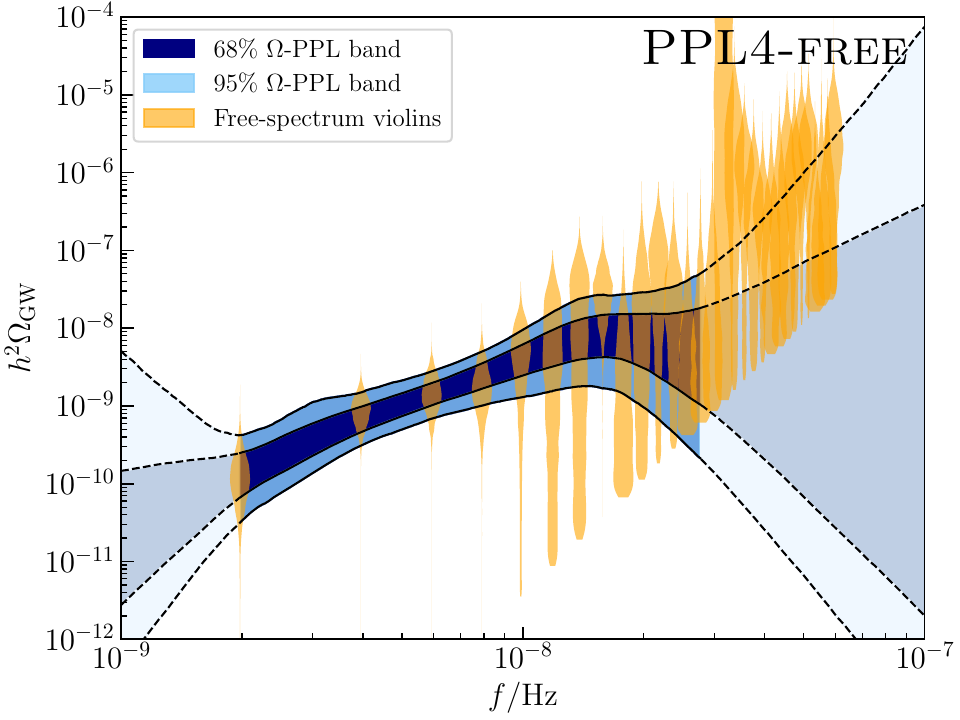}\quad
\includegraphics[width=0.23\textwidth]{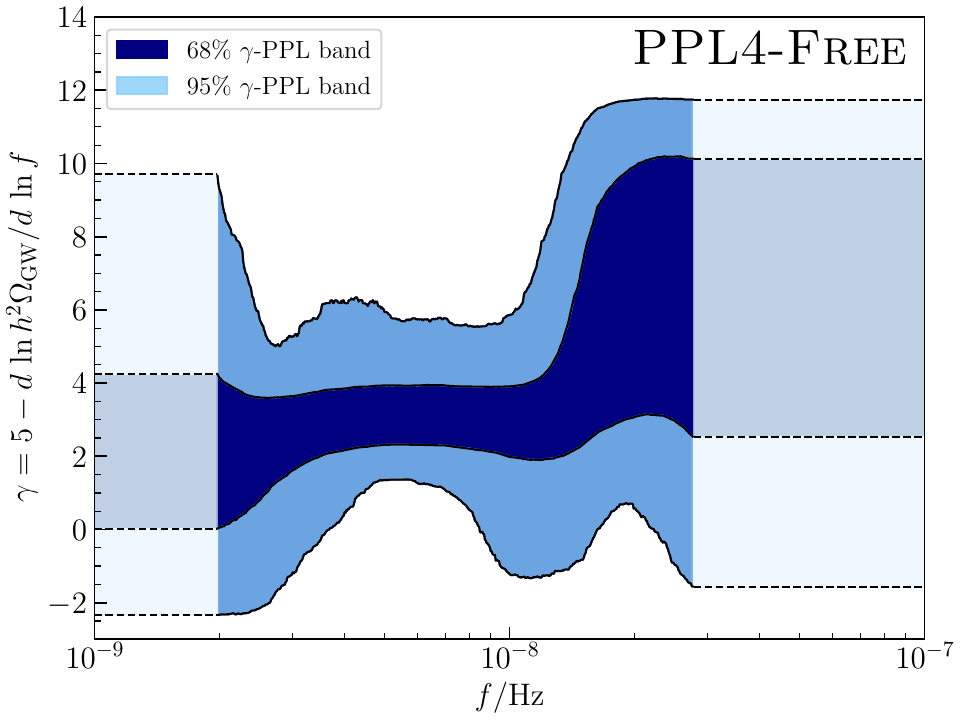}

\includegraphics[width=0.23\textwidth]{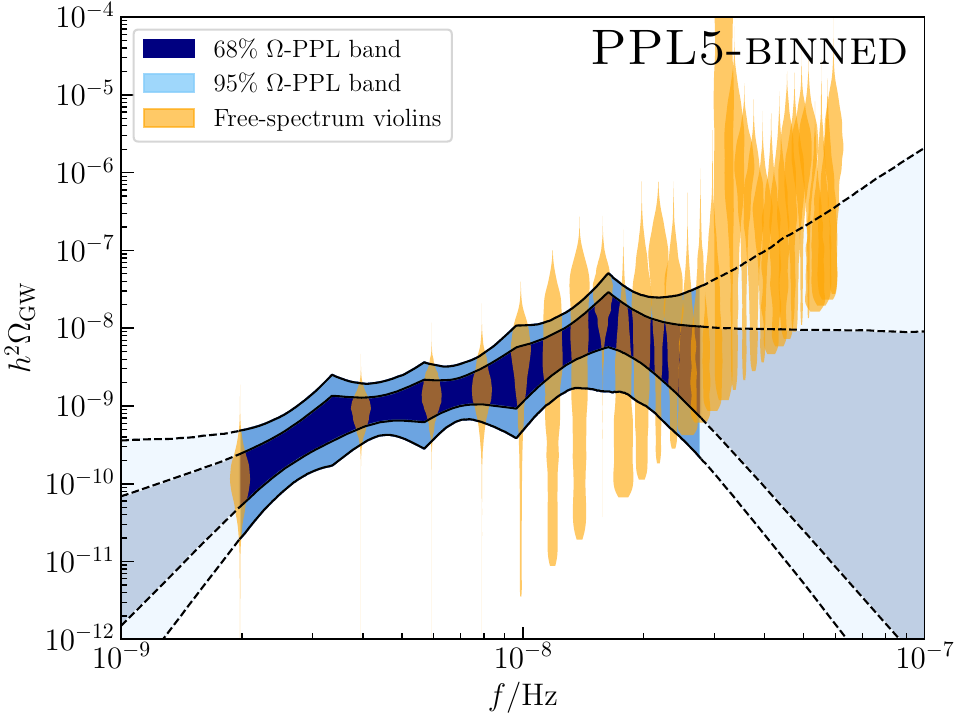}\quad
\includegraphics[width=0.23\textwidth]{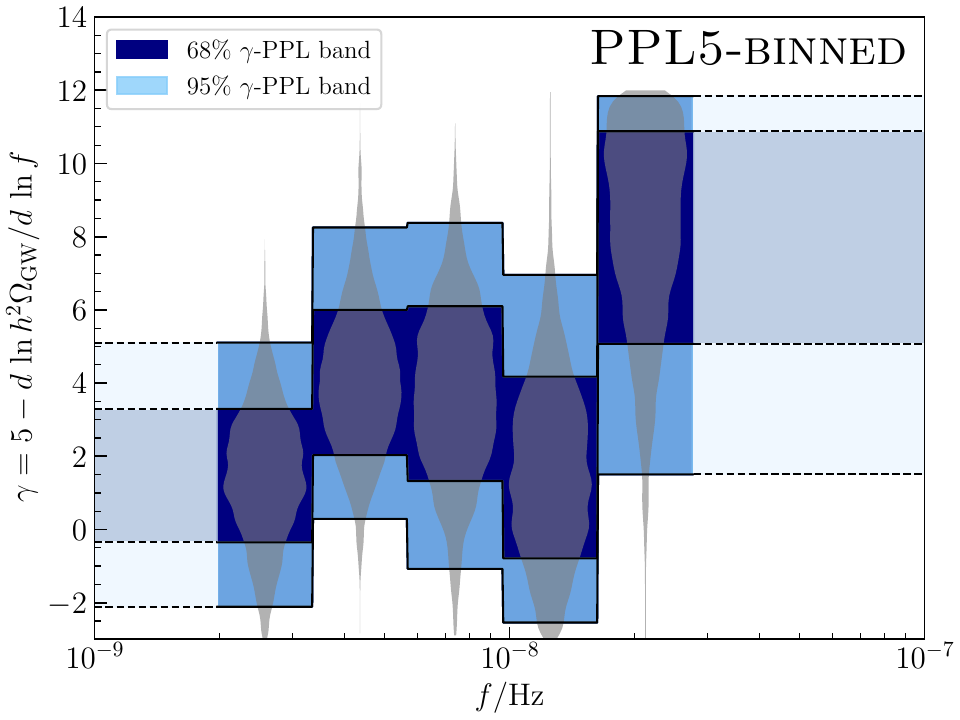} \quad
\includegraphics[width=0.23\textwidth]{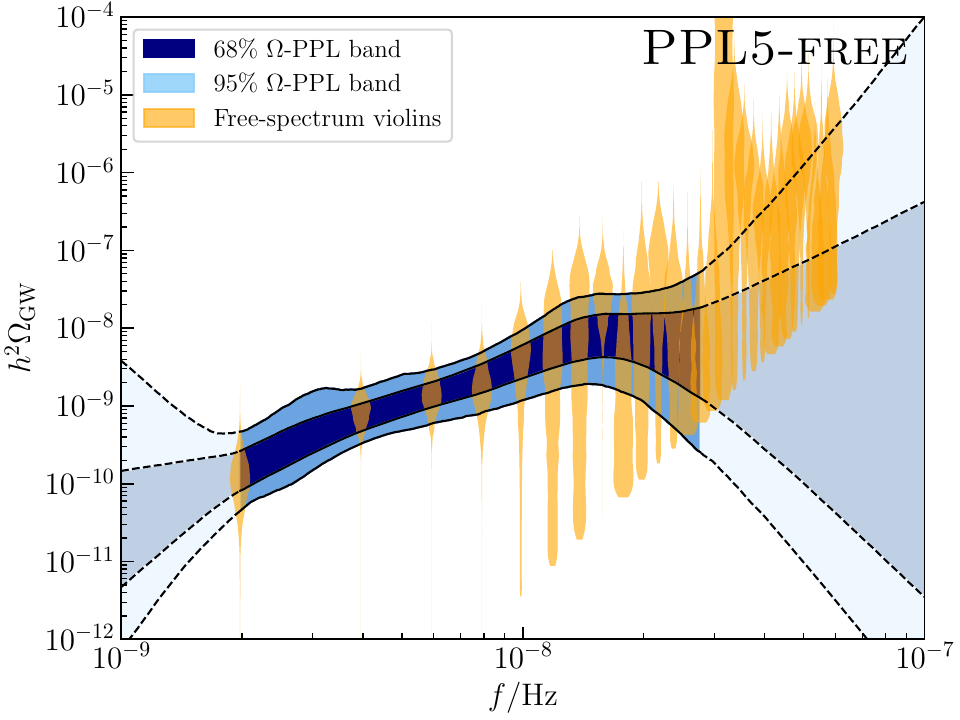}\quad
\includegraphics[width=0.23\textwidth]{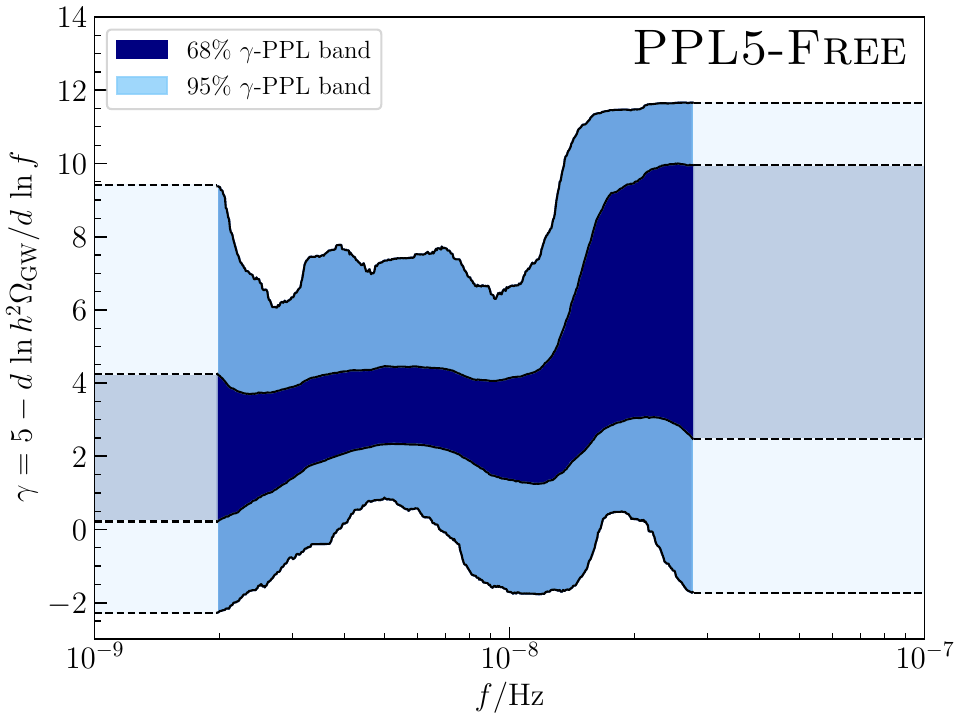}

\includegraphics[width=0.23\textwidth]{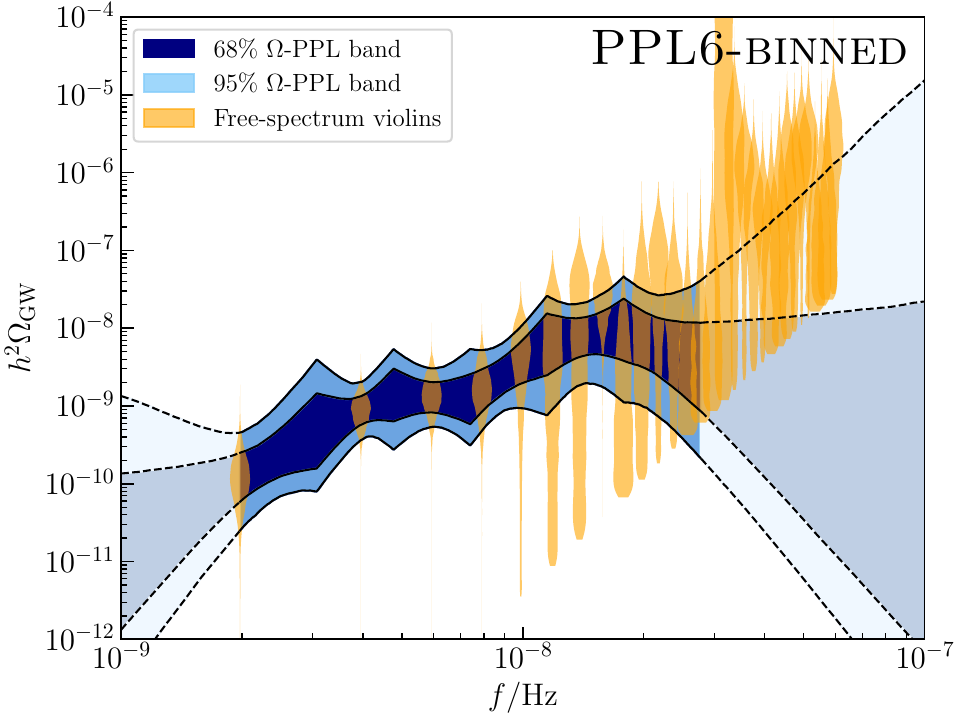}\quad
\includegraphics[width=0.23\textwidth]{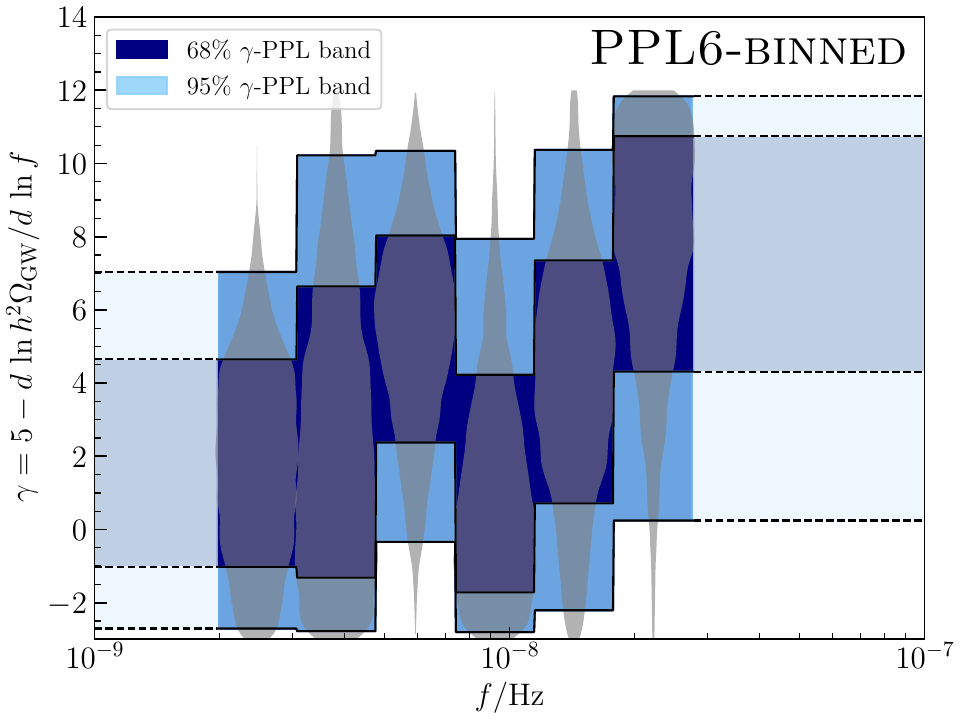} \quad
\includegraphics[width=0.23\textwidth]{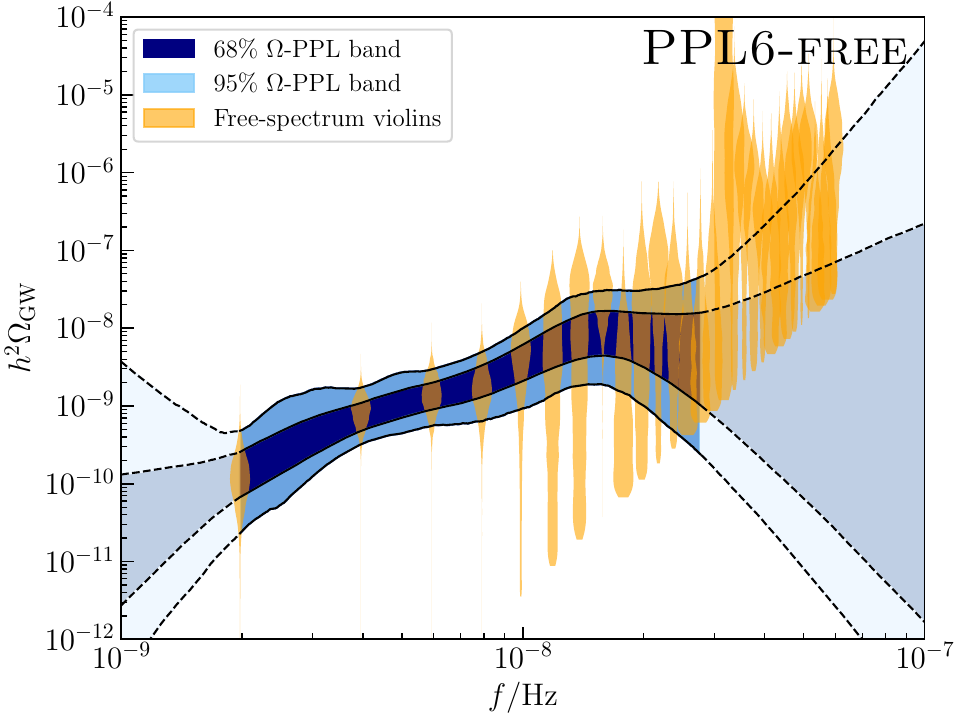}\quad
\includegraphics[width=0.23\textwidth]{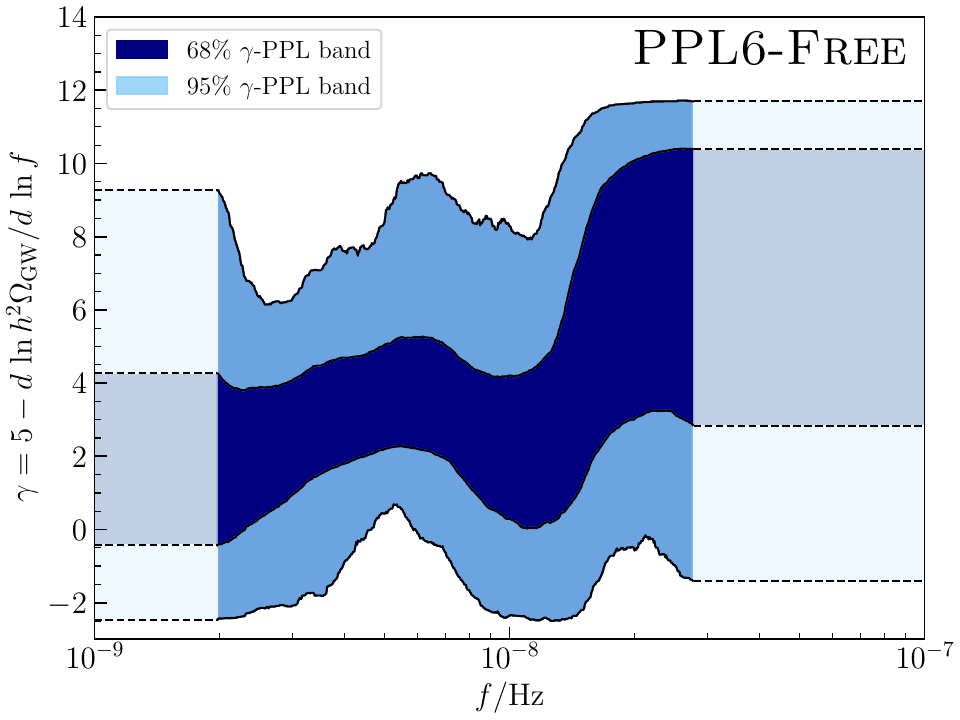}
\end{center}
\caption{$\Omega$-PPL and $\gamma$-PPL bands for all individual \textsc{binned} and \textsc{free} models considered in this work (see text for details).}
\label{fig:PPLnbands}

\end{figure*}


\subsection{Results and discussion: Individual models}

\begin{figure*}
\begin{center}
\includegraphics[width=0.4\textwidth]{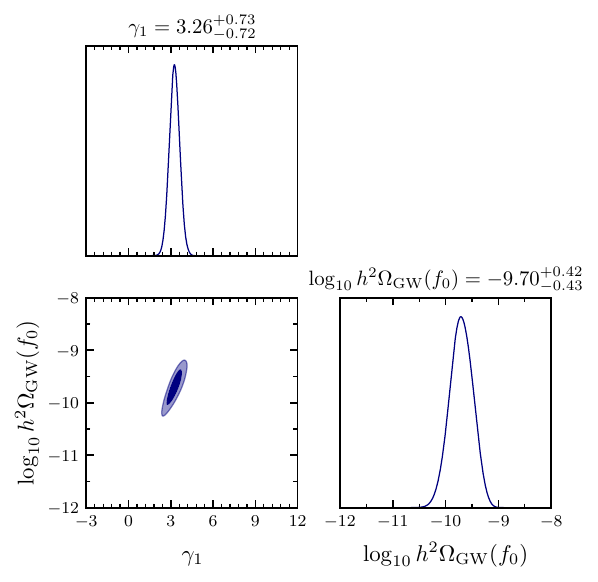}\qquad
\includegraphics[width=0.4\textwidth]{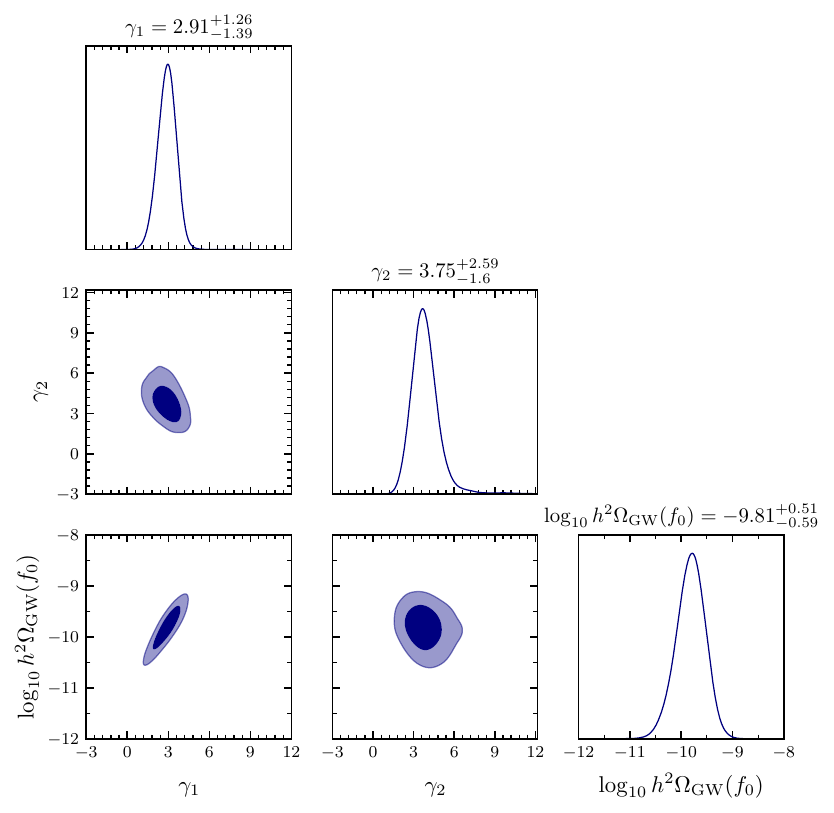}

\includegraphics[width=0.4\textwidth]{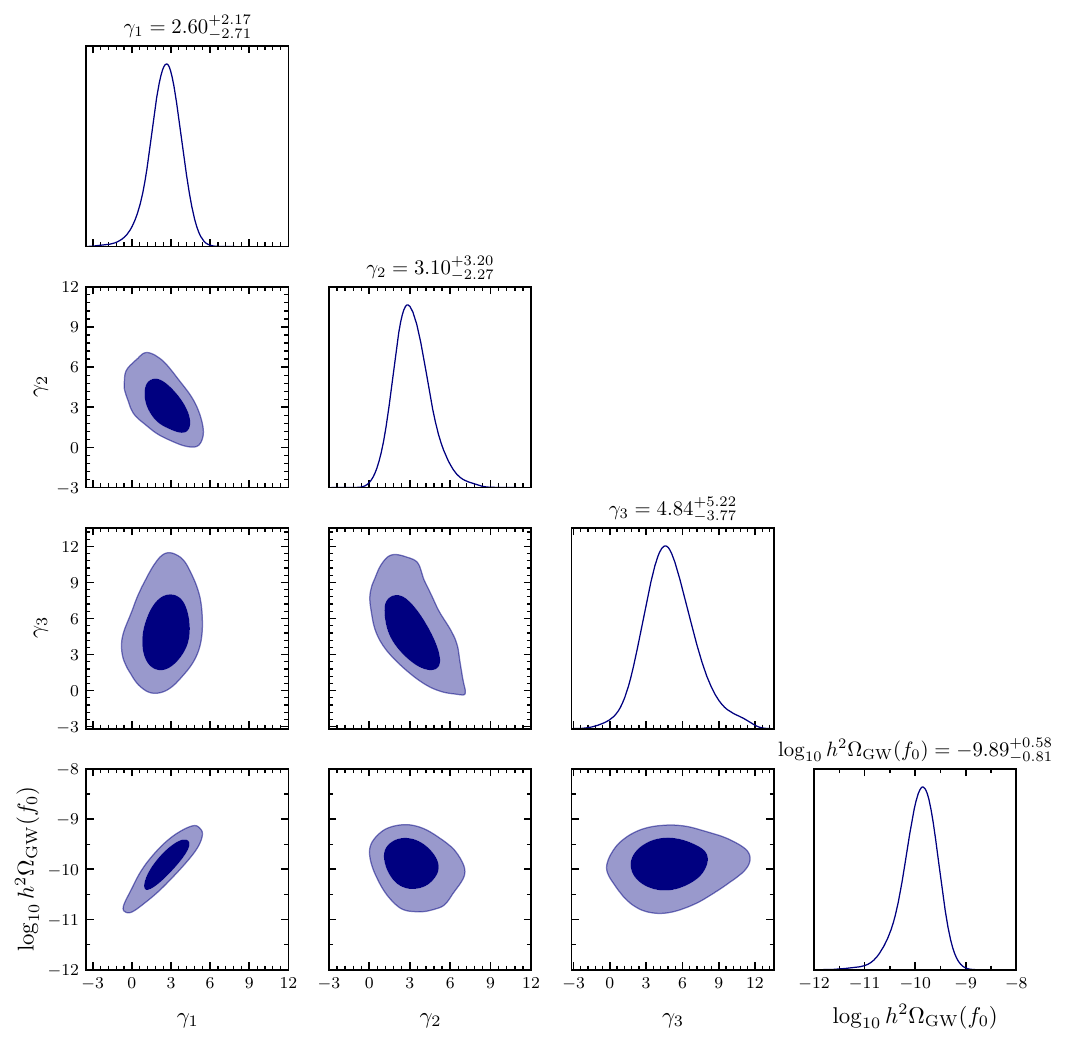}\qquad
\includegraphics[width=0.4\textwidth]{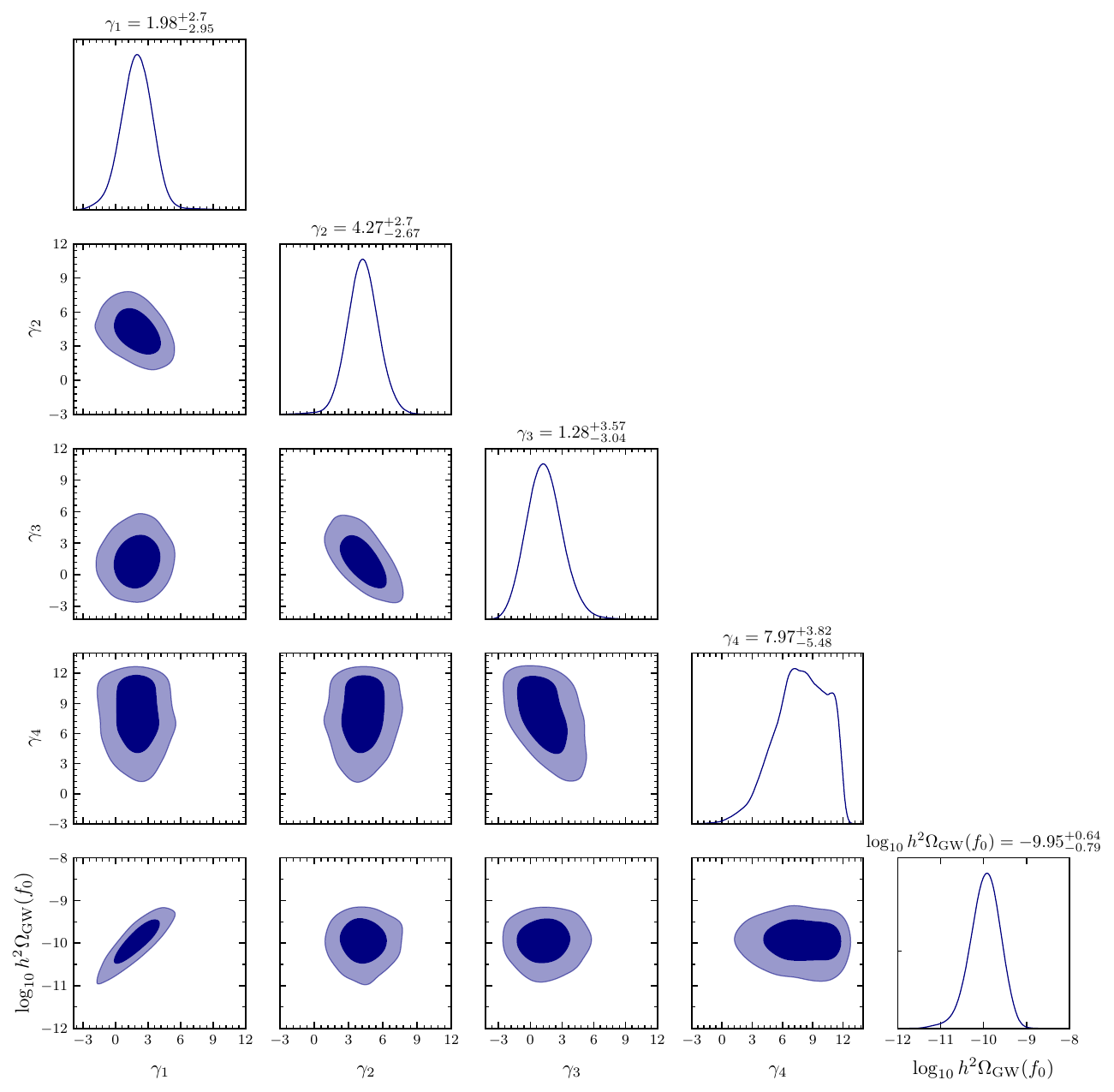}

\includegraphics[width=0.4\textwidth]{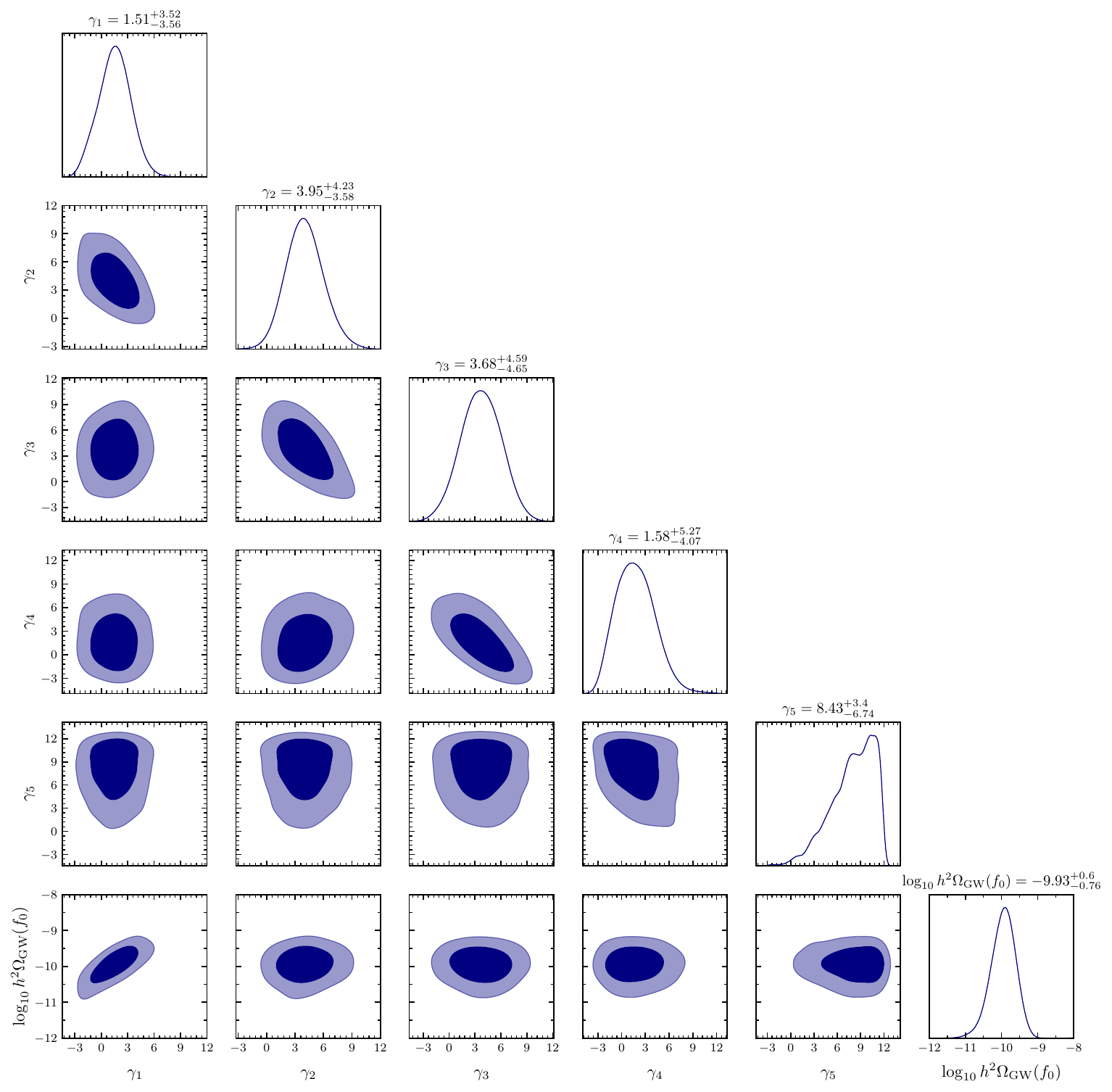}\qquad
\includegraphics[width=0.4\textwidth]{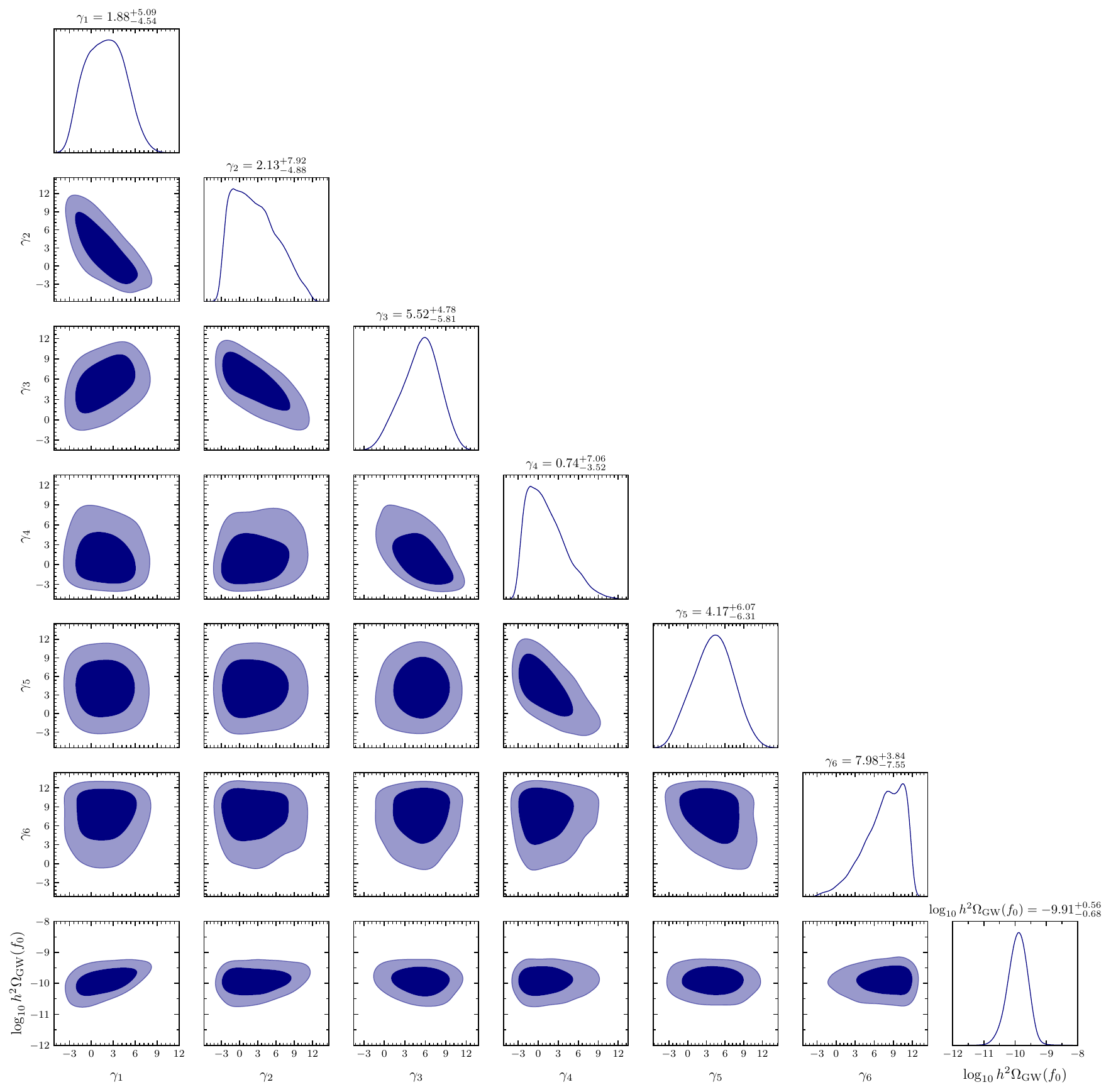}
\end{center}
\caption{Corner plots for the \textsc{binned} models. Parameter regions shaded in darker and lighter colors correspond to $68\%$ and $95\%$ credible regions, respectively. At the top of each column, we state median values and 95\% equal-tailed credible intervals.}
\label{fig:corner-binned}
\end{figure*}

\begin{figure*}
\begin{center}
\includegraphics[width=0.4\textwidth]{fig/PPL1_Corner.pdf}\qquad
\includegraphics[width=0.4\textwidth]{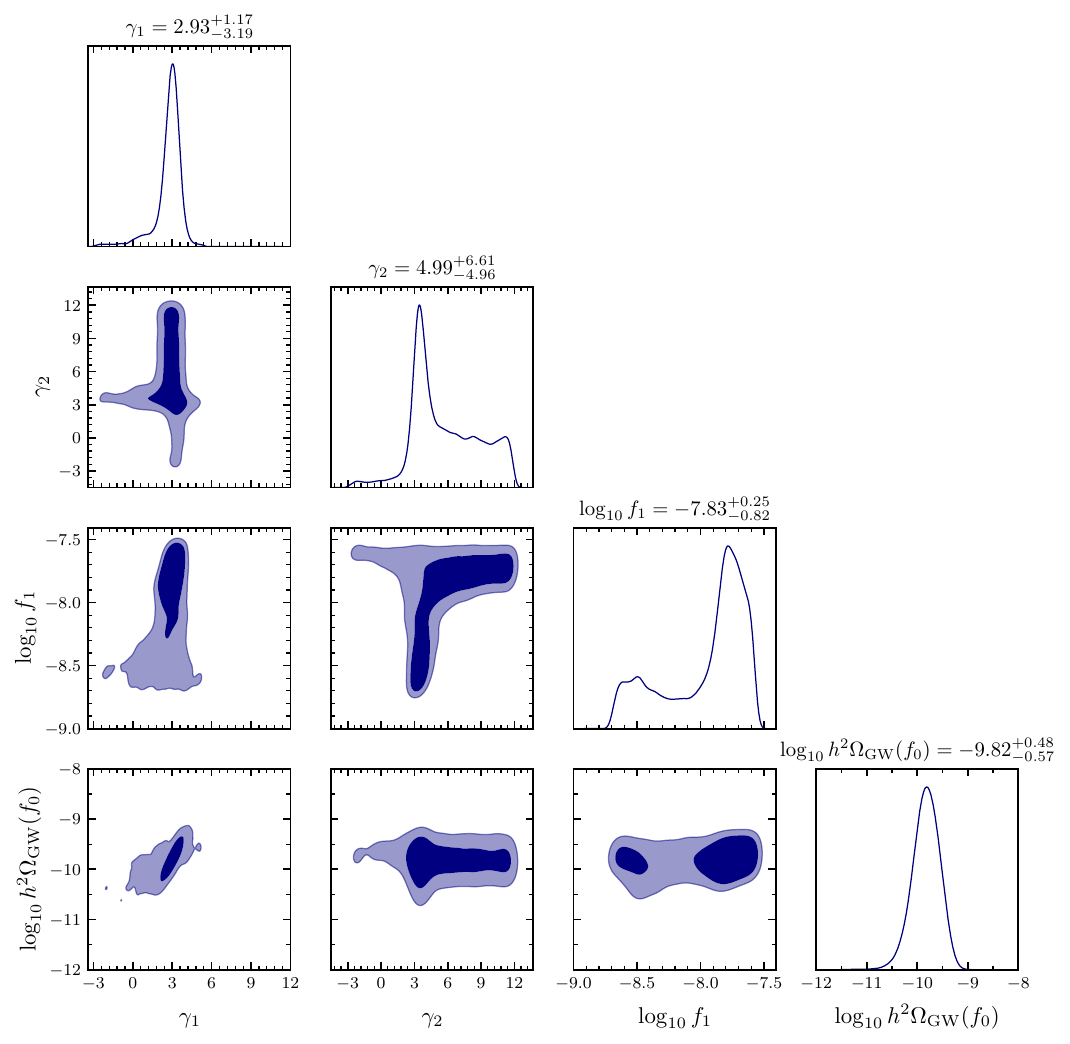}

\includegraphics[width=0.4\textwidth]{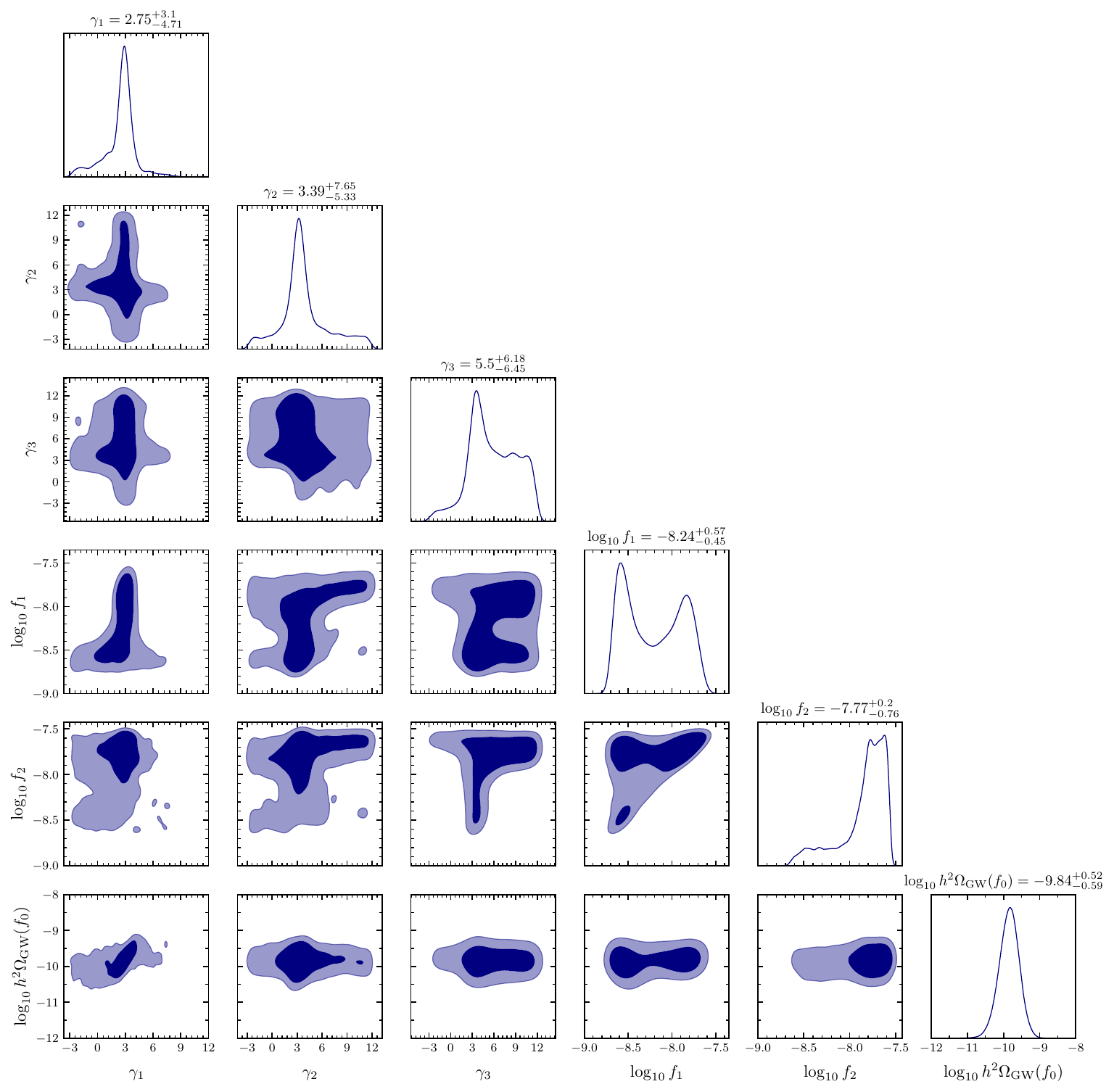}\qquad
\includegraphics[width=0.4\textwidth]{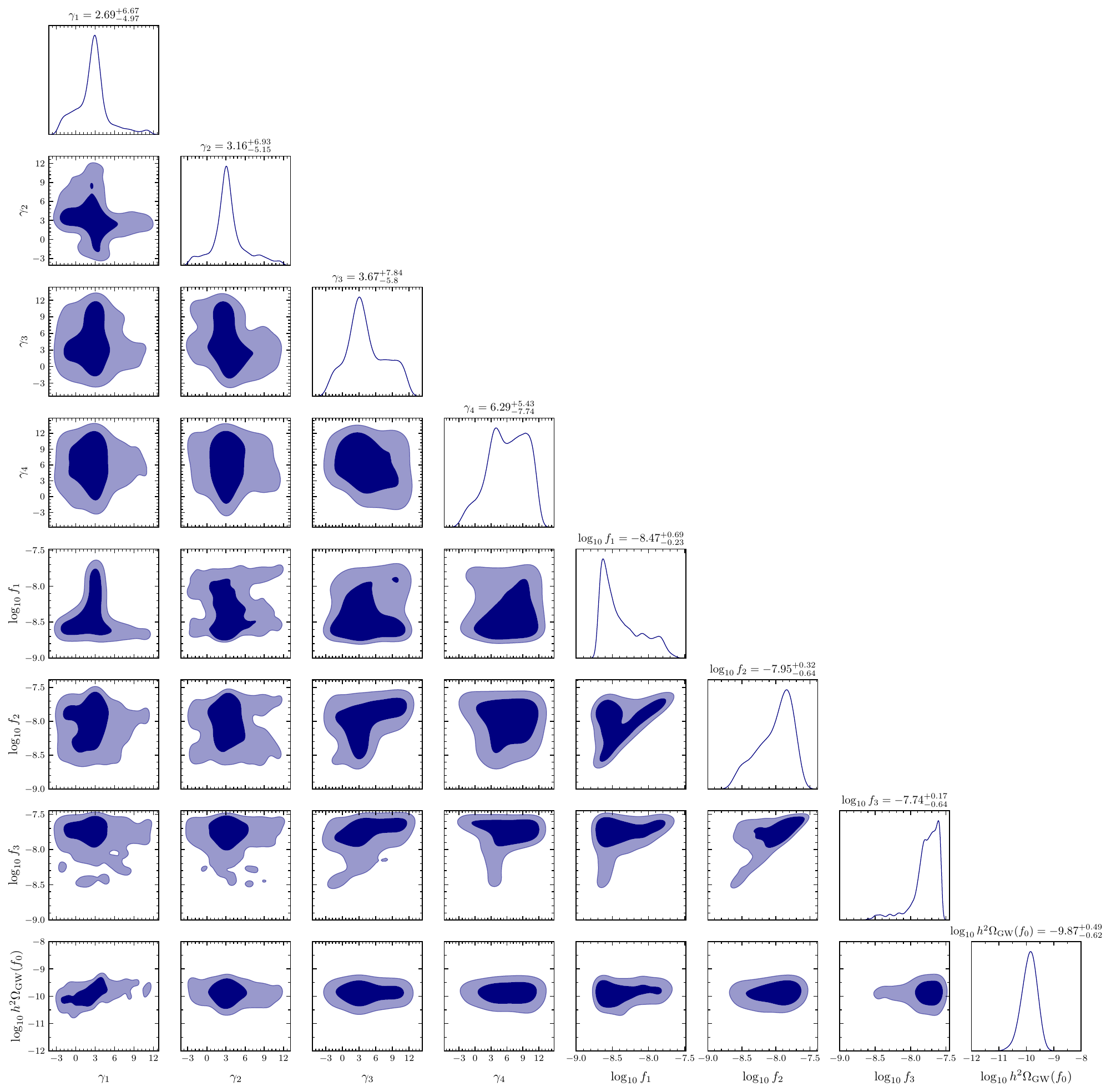}

\includegraphics[width=0.4\textwidth]{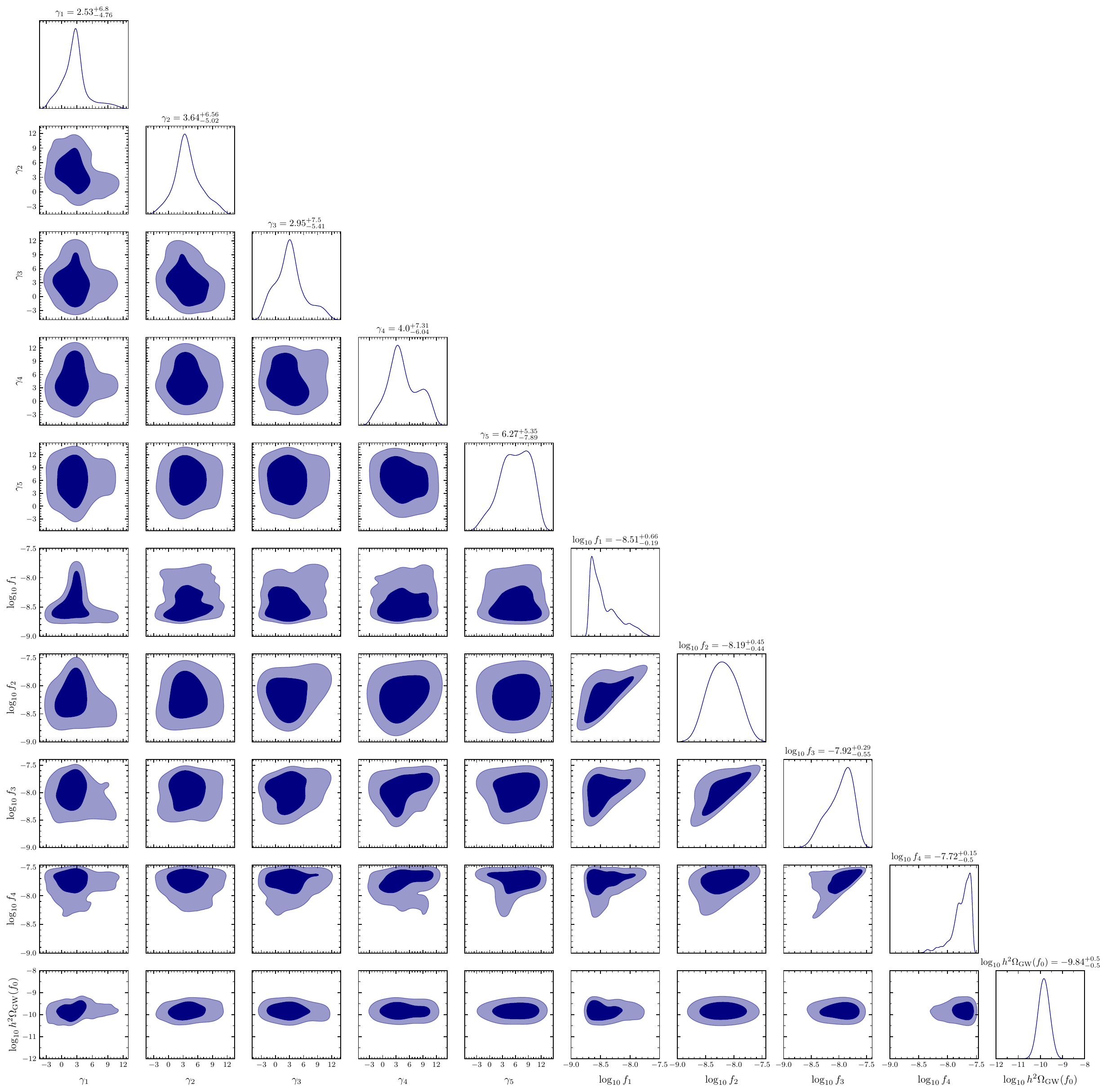}\qquad
\includegraphics[width=0.4\textwidth]{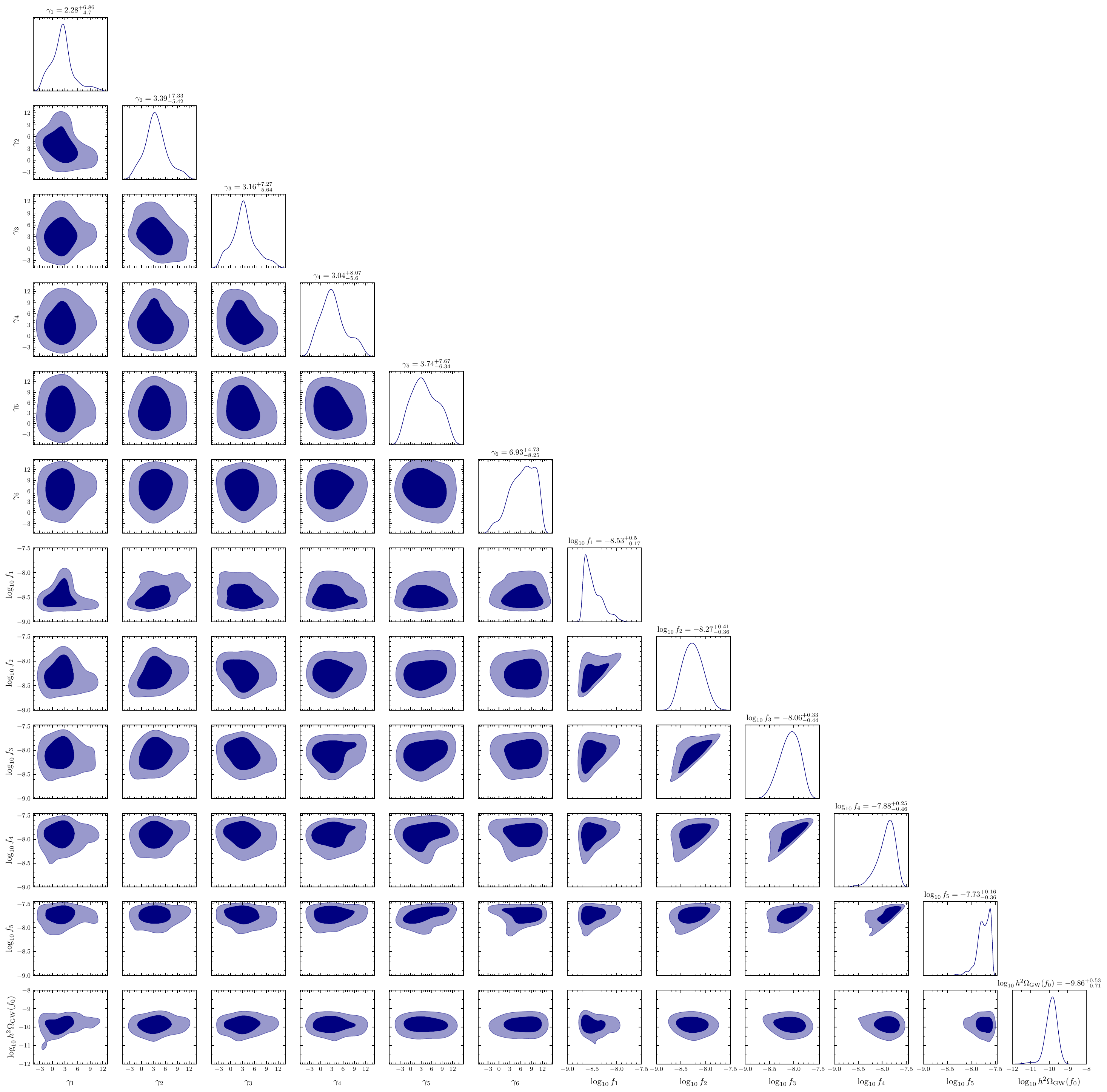}
\end{center}
\caption{Corner plots for the \textsc{free} models. Parameter regions shaded in darker and lighter colors correspond to $68\%$ and $95\%$ credible regions, respectively. At the top of each column, we state median values and 95\% equal-tailed credible intervals.}
\label{fig:corner-free}
\end{figure*}

The results of our Bayesian fits at the level of individual PPL models are shown in Figs.~\ref{fig:PPLnbands}, \ref{fig:corner-binned}, and \ref{fig:corner-free}. In Fig.~\ref{fig:PPLnbands}, we present what we refer to as the $\Omega$-PPL and $\gamma$-PPL bands for all of our \textsc{binned} and \textsc{free} models. Slicing these bands at a given frequency, one obtains the distributions of $h^2\Omega_{\rm GW}$ and $\gamma$ values at this frequency that are induced by the posterior density in the GWB parameter space. In practice, we construct all $\Omega$-PPL and $\gamma$-PPL bands from our MCMC chains as follows: For a given model, we first pick a large number of MCMC samples. Each sample corresponds to a choice of GWB parameters, which in turn fix the precise form of the GWB spectrum. From this spectrum, we then read off the $h^2\Omega_{\rm GW}$ and $\gamma$ values at any given frequency. Repeating this procedure for all selected MCMC samples thus results in distributions of $h^2\Omega_{\rm GW}$ and $\gamma$ values at all frequencies of interest. The $68\%$ and $95\%$ bands in Fig.~\ref{fig:PPLnbands} specifically mark the boundaries of the median credible intervals of $68\%$ and $95\%$ probability in these distributions at each frequency.

In our description of the NG15 timing residuals, the GWB component is only present at frequencies between $f_{\rm min} = \sfrac{1}{T_{\rm obs}}$ and $f_{\rm max} = \sfrac{14}{T_{\rm obs}}$. Correspondingly, the $\Omega$-PPL and $\gamma$-PPL bands are only well defined in this frequency range, which we indicate by darker colors in Fig.~\ref{fig:PPLnbands}. Meanwhile, at smaller and larger frequencies, we extrapolate our results, which we indicate by lighter colors in Fig.~\ref{fig:PPLnbands}. In practice, we construct these parts of our $\Omega$-PPL and $\gamma$-PPL bands by a slight extension of our spectral models, assuming that the GWB spectrum maintains its power-law shape with spectral index $\gamma_1$ even below $f_{\rm min}$ and its power-law shape with spectral index $\gamma_n$ even above $f_{\rm max}$. In the case of the $\Omega$-PPL bands, the resulting extrapolation highlights the behavior of the bands around $f_{\rm min}$ and $f_{\rm max}$ and the trend in the bands beyond these frequencies. In the case of the $\gamma$-PPL bands, on the other hand, the resulting extrapolation simply reflects our underlying assumption, namely, that the spectral index remains fixed at $\gamma_1$ for $f < f_{\rm min}$ as well as fixed at $\gamma_n$ for $f > f_{\rm max}$. In all cases, the extrapolated parts of the $\Omega$-PPL and $\gamma$-PPL bands serve as a useful visual aid. However, beyond that, they contain no new information, which is why we include them in Fig.~\ref{fig:PPLnbands} for purely illustrative purposes. 

Next, let us compare our results for the \textsc{binned} and \textsc{free} models. In the case of our simplest model, PPL1, we find identical results by construction. As mentioned before, \textsc{PPL1-binned} and \textsc{PPL1-free} denote one and the same model, simply because PPL1 features no node frequency that could either remain fixed or vary freely. In fact, our results for PPL1 are in one-to-one correspondence with the results for the CPL fit to the NG15 data presented in \cite{NANOGrav:2023gor}, where this model is referred to as the HD$^\gamma$ model. The $\Omega$-PPL band for PPL1 in Fig.~\ref{fig:PPLnbands} contains the same information as the blue band labeled ``power-law posterior'' in Fig.~1(a) of \cite{NANOGrav:2023gor}, and the $\gamma$-PPL band for PPL1 in Fig.~\ref{fig:PPLnbands} represents the same posterior density as the $\gamma_{\rm GWB}$ posterior density in Fig.~1(b) of \cite{NANOGrav:2023gor}.

For all other PPL models, the results of the \textsc{binned} models differ from those of the \textsc{free} models. In the case of the \textsc{binned} models, the $\Omega$-PPL and $\gamma$-PPL bands in Fig.~\ref{fig:PPLnbands} clearly reflect the fixed frequency bins entering the construction of the models, which always remain the same, no matter the choice of GWB model parameters. In the case of the \textsc{free} models, on the other hand, all boundaries between frequency bins are washed out, as the internal node frequencies allowed to move around. Despite these differences, it is apparent from Fig.~\ref{fig:PPLnbands} that the $\Omega$-PPL and $\gamma$-PPL bands for the \textsc{binned} models represent reasonable (literally, binned) approximations of their respective counterparts in the \textsc{free} models.

All PPL reconstructions of the GWB spectrum are also consistent with the NG15 violins, which we show on top of each $\Omega$-PPL band in Fig.~\ref{fig:PPLnbands}. In particular, one can ignore the fact that some $\Omega$-PPL bands extend to lower amplitudes than the NG15 violins around $f_{\rm max}$; the NG15 violins are simply cut off from below by a prior bound that rises like $f^5$ in units of $h^2\Omega_{\rm GW}$. 

The $\gamma$-PPL bands for the \textsc{binned} models allow us to construct what may be called $\gamma$ violins; see the gray violins in Fig.~\ref{fig:PPLnbands}. These $\gamma$ violins represent the marginalized posterior densities for the spectral indices $\gamma_i$ in the \textsc{binned} models. In a sense, these violins can thus be regarded as the equivalent\,---\,at the level of the index rather than the amplitude of the GWB spectrum\,---\,of the NG15 violins in the free spectral model. Furthermore, we observe that, at frequencies close to $f_{\rm max}$, the $\gamma$-PPL bands in both the \textsc{binned} and \textsc{free} models often saturate the upper boundary of the $\gamma_i$ prior range (see Tab.~\ref{tab:priors}). This observation reflects the lack of constraining power of the NG15 data at larger frequencies, which is consistent with the behavior of the $\Omega$-PPL bands and the NG15 violins at larger frequencies. 


The posterior densities for the actual GWB model parameters, i.e., $\left\{h^2\Omega_{\rm GW}\left(f_0\right),\gamma_1,\cdots,\gamma_n\right\}$ in the \textsc{binned} models and $\left\{h^2\Omega_{\rm GW}\left(f_0\right),\gamma_1,\cdots,\gamma_n, f_1,\cdots,f_{n-1}\right\}$ in the \textsc{free} models, can be found in Figs.~\ref{fig:corner-binned} and \ref{fig:corner-free}. These figures present corner plots, featuring one-dimensional (1D) marginalized posterior densities in the plots on the diagonal as well as 2D marginalized posterior densities in the off-diagonal plots, for all PPL models of interest. Again, the corner plots for \textsc{PPL1-free} and \textsc{PPL1-binned} are identical by definition. For all other models, we notice several differences.

The posterior densities in the \textsc{binned} models are all well converged and mostly unimodal. In particular, no 2D marginalized posterior density exhibits a flat direction that could in principle be continued far beyond the boundaries of our prior ranges. The posterior densities of the \textsc{binned} models are therefore well behaved, which underlines the utility of these models, despite the fact that we merely regard them as an intermediate step in the construction of the \textsc{free} models. The situation in the \textsc{free} models is more complicated. For these models, multimodal distributions are more common, and parameter degeneracies often result in flat directions that extend all the way up to the boundaries of the prior ranges. Nonetheless, the 1D marginalized posterior densities in Fig.~\ref{fig:corner-free} indicate that our MCMC chains cover the most relevant regions of the GWB parameter space, which justifies our prior choices in Tab.~\ref{tab:priors}. Finally, we note that the 2D marginalized posterior densities for pairs of node frequencies (in all \textsc{free} models with at least two node frequencies) reflect the correct ordering of the internal node frequencies, $f_1 \leq f_2 \leq \cdots \leq f_{n-1}$. In the \textsc{PPL3-free} model, e.g., the 2D marginalized posterior density for $\log_{10}f_1$ and $\log_{10} f_2$ is only nonzero when the two frequencies are correctly ordered, $f_1 \leq f_2$. 


\subsection{Bayesian model comparison and average}

Our central goal in this paper is to perform a Bayesian average over the results for the individual PPL models. To this end, we need to carry out a Bayesian model comparison that allows us to determine appropriate weights (probabilities) $P_n = P(\textrm{PPL$n$}|D)$ for each PPL model. According to Bayes' theorem, these model probabilities can be expressed in terms of marginal likelihoods (model evidences) $\mathcal{L}(D|\textrm{PPL$n$})$ and model priors $\pi(\textrm{PPL$n$})$, 
\begin{equation}
P_n = \frac{\mathcal{L}(D|\textrm{PPL$n$})\,\pi(\textrm{PPL$n$})}{\sum_m \mathcal{L}(D|\textrm{PPL$m$})\,\pi(\textrm{PPL$m$})} \,.
\end{equation}
Then, if we assume equal prior probability for all models, $P_n$ can be written as a normalized marginal likelihood,  
\begin{equation}
P_n = \frac{\mathcal{L}(D|\textrm{PPL$n$})}{\sum_m \mathcal{L}(D|\textrm{PPL$m$})} \,,
\end{equation}
which is sometimes referred to as a relative evidence. Next, we can divide by the marginal likelihood of the \textsc{PPL1} model and express $P_n$ in terms of Bayes factors, 
\begin{equation}
\label{eq:PnBn1}
P_n = \frac{\mathcal{B}_{n1}}{\sum_m \mathcal{B}_{m1}} \,, \qquad  \mathcal{B}_{n1} = \frac{\mathcal{L}(D|\textrm{PPL$n$})}{\mathcal{L}(D|\textrm{PPL1})} \,.
\end{equation}

In the following, we shall use product-space sampling methods~\citep{10.2307/2346151,10.2307/1391010,Hee:2015eba} in order to compute Bayes factors. Product-space sampling, however, works best for similar models that provide fits of the data of comparable quality. Therefore, instead of evaluating the Bayes factors $\mathcal{B}_{n1}$ directly, we first rewrite them in two steps that prove advantageous from the perspective of our numerical analysis. First, we make use of the fact that Bayes factors are transitive, which means that $\mathcal{B}_{n1}$ can be written as a product of Bayes factors $\mathcal{B}_{k\,k-1}$ for pairs of successive PPL models,
\begin{equation}
\label{eq:Bn1prod}
\mathcal{B}_{n1} = \prod_{k=2}^n \mathcal{B}_{k\,k-1} \,.
\end{equation}
The \textsc{binned} and \textsc{free} PPL$k$ and PPL$k-1$ models differ by only one and two parameters, respectively, which facilitates the evaluation of the Bayes factors $\mathcal{B}_{k\,k-1}$. However, we still do not compute $\mathcal{B}_{k\,k-1}$ directly. In a second step, we distinguish between HD and CURN versions of the PPL$k$ and PPL$k-1$ models and write
\begin{equation}
\mathcal{B}_{k\,k-1} =  \frac{\mathcal{B}_k^{\rm HD/CURN}}{\mathcal{B}_{k-1}^{\rm HD/CURN}}\:\mathcal{B}_{k\,k-1}^{\rm CURN} \,.
\end{equation}
Here, $\mathcal{B}_{k\,k-1}$ assumes HD cross-correlations in Eq.~\eqref{eq:SGWB} for both models, PPL$k$ and PPL$k-1$, whereas $\mathcal{B}_{k\,k-1}^{\rm CURN}$ assumes no cross-correlations ($\Gamma_{ab}\rightarrow \delta_{ab}$) in Eq.~\eqref{eq:SGWB}, 
\begin{equation}
\mathcal{B}_{k\,k-1}^{\rm CURN} =  \frac{\mathcal{L}(D|\textrm{PPL$k$, CURN})}{\mathcal{L}(D|\textrm{PPL$k-1$, CURN})} \,.
\end{equation}
In that sense, $\mathcal{B}_{k\,k-1}^{\rm CURN}$ is the CURN analog of the Bayes factor that we are actually interested in, $\mathcal{B}_{k\,k-1}$. Meanwhile, $\mathcal{B}_k^{\rm HD/CURN}$ compares the marginal likelihoods of PPL$k$ in its HD and CURN versions, respectively, 
\begin{equation}
\mathcal{B}_k^{\rm HD/CURN} = \frac{\mathcal{L}(D|\textrm{PPL$k$, HD})}{\mathcal{L}(D|\textrm{PPL$k$, CURN})} \,.
\end{equation}

With these ingredients at our disposal, we are now ready to apply product-space sampling, which effectively turns model selection into a parameter estimation problem. Specifically, in order to compute $\mathcal{B}_{k\,k-1}^{\rm CURN}$, we sample the union of unique parameters of PPL$k$ and PPL$k-1$ in their CURN versions, alongside an indexing parameter that activates the selected parameter space and likelihood model. $\mathcal{B}_{k\,k-1}^{\rm CURN}$ is then given by the ratio between the number of samples collected for each model. Furthermore, we make use of likelihood reweighting~\citep{Hourihane:2022ner} in order to evaluate $\mathcal{B}_k^{\rm HD/CURN}$. In principle, there are two use cases for likelihood reweighting: this technique allows one to model a target posterior by reweighting an approximate posterior, and it yields an estimate of the corresponding Bayes factor. In our case, we shall use likelihood reweighting solely for the latter purpose, i.e., for computing $\mathcal{B}_k^{\rm HD/CURN}$. To do so, we evaluate the likelihood of PPL$k$ in its HD version on the accepted samples for PPL$k$ in its CURN version that we obtain from our product-space sampling runs. For each sample, we thus obtain a weight factor, 
\begin{equation}
w_i =  \frac{\mathcal{L}(D|\bm{\theta}_i, \textrm{PPL$k$, HD})}{\mathcal{L}(D|\bm{\theta}_i, \textrm{PPL$k$, CURN})} \,.
\end{equation}
In the CURN versions of the PPL models, we assume the same prior densities as in their HD versions, which allows us to compute $\mathcal{B}_k^{\rm HD/CURN}$ as the sample mean of the weight factors $w_i$~\citep{Hourihane:2022ner}. We stress once more that this is the only part of our analysis in which we resort to  likelihood reweighting. All other results for the PPL models in their HD version, in particular all results in Fig.~\ref{fig:PPLnbands}, \ref{fig:corner-binned}, and \ref{fig:corner-free}, are based on Bayesian MCMC fits employing the full HD likelihoods. 


\begin{figure}
\begin{center}
\includegraphics[width=0.47\textwidth]{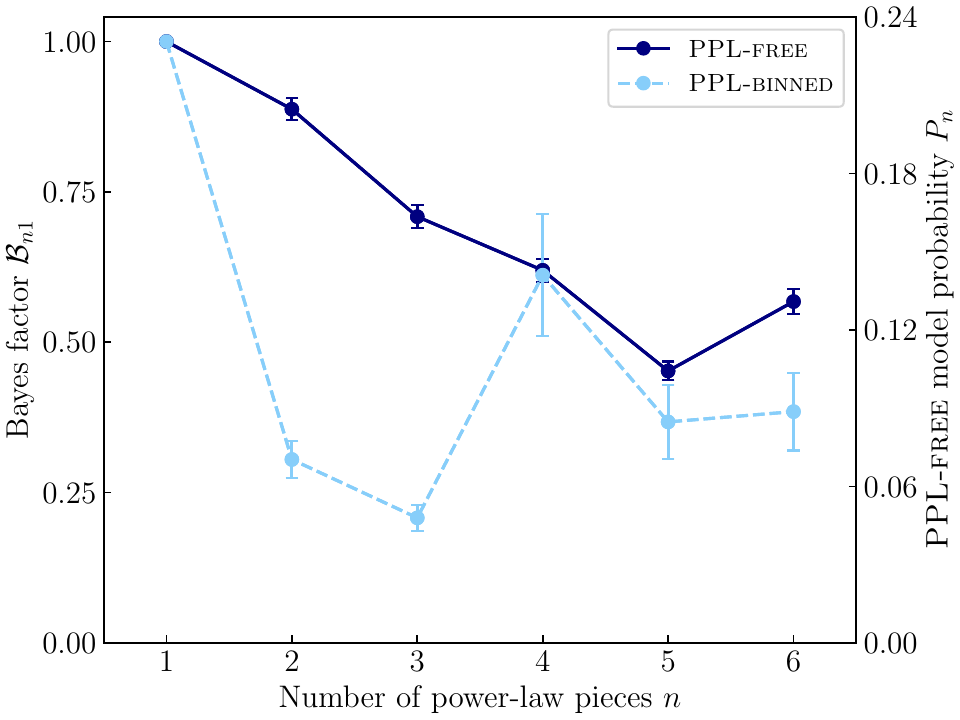}
\end{center}
\caption{Bayes factors $\mathcal{B}_{n1}$ for the model comparisons PPL$n$ versus PPL$1$, both for the \textsc{binned} and \textsc{free} models. Normalizing the $\mathcal{B}_{n1}$ by their sum yields the model probabilities $P_n$ [see Eq.~\eqref{eq:PnBn1}, shown here for the \textsc{free} models].}
\label{fig:bf}
\end{figure}


\begin{figure*}
\begin{center}
\includegraphics[width=0.49\textwidth]{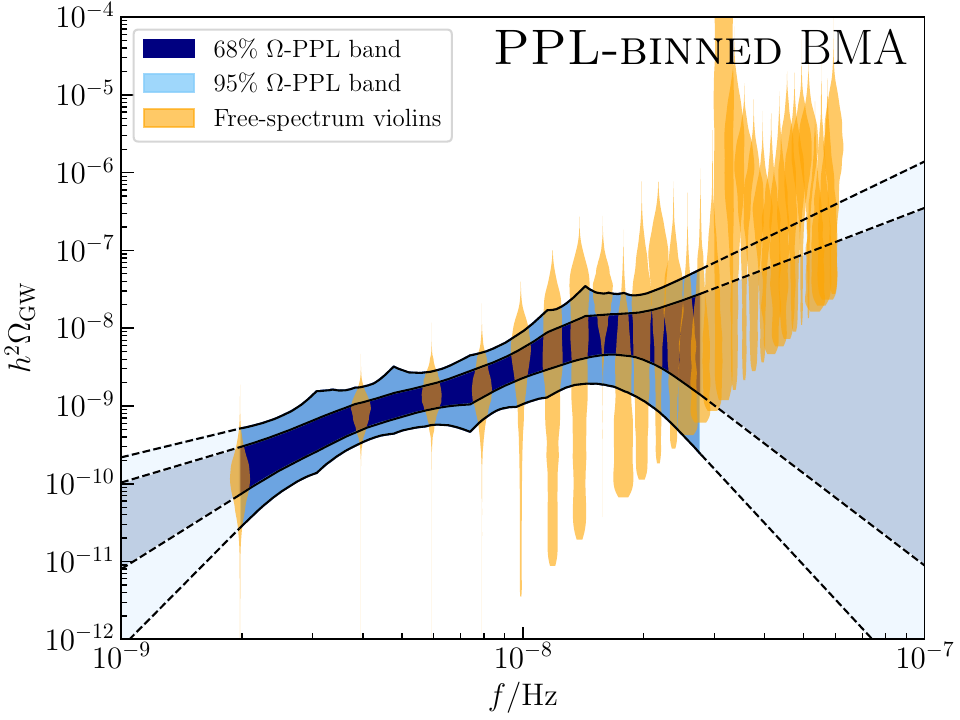}\quad
\includegraphics[width=0.49\textwidth]{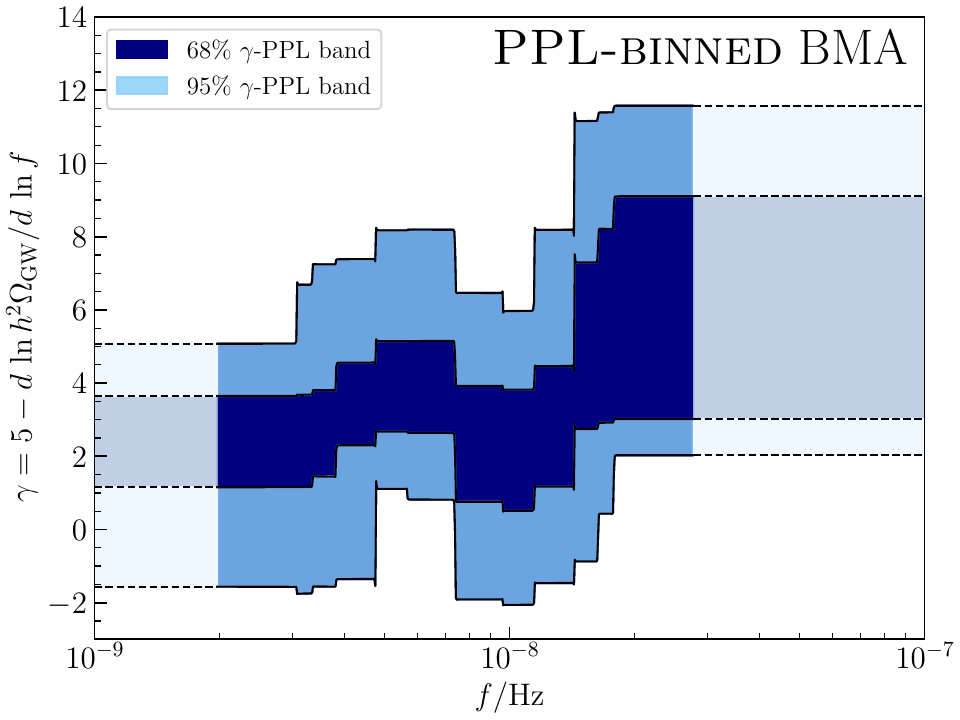}
\end{center}
\caption{\textit{Left panel:} PPL reconstruction of the energy density power spectrum $h^2\Omega_{\rm GW}$ for the GWB signal in the NG15 data, based on the BMA of six PPL models with a fixed number of equidistant internal nodes (\textsc{PPL1-binned},\,...\,\textsc{PPL6-binned}). \textit{Right panel:} Same, but for the frequency-dependent spectral index of the energy density power spectrum, $\gamma = 5- d \ln h^2\Omega_{\rm GW}/ d\ln f$. Both reconstructions cover the frequency range from $F_1 = \sfrac{1}{T_{\rm obs}}$ to $F_{14} = \sfrac{14}{T_{\rm obs}}$, where $T_{\rm obs} = 16.03\,\textrm{yr}$ (see text for details). In the left panel, we also show the NG15 violins in orange, which represent the amplitude posterior densities for the free spectral model. The lower end of the violins at higher frequencies follows from the prior density chosen in~\cite{NANOGrav:2023gor}.}
\label{fig:og-binned-bma}
\end{figure*}


\subsection{Results and discussion: Averaged models}

Our results for the Bayes factors $\mathcal{B}_{n1}$ and model probabilities $P_n$ are shown in Fig.~\ref{fig:bf}. The central values and error bars in this plot correspond to mean values and standard deviations that we compute using statistical bootstrapping methods~\citep{Efron:1986hys}. As evident from Fig.~\ref{fig:bf}, the NG15 data is best fitted by the PPL1 model: adding more complexity by introducing further power-law segments results in a decreasing Bayes factor $\mathcal{B}_{n1}$. This observation reflects our expectation and justifies the construction of BMAs over a finite a number of individual PPL models. In the case of the \textsc{binned} models, we, moreover, observe that the data disfavors power-law breaks at the fixed node frequencies $f_1$ and $f_2$. If these frequencies are allowed to vary, the corresponding Bayes factors $\mathcal{B}_{21}$ and $\mathcal{B}_{31}$ are less suppressed. The gap between the Bayes factors for the \textsc{binned} and \textsc{free} models only partially closes when a third internal node frequency, $f_3$, is added. That is, the \textsc{binned} models can only approximately match the flexibility of their \textsc{free} counterparts starting from PPL$4$. At the same time, we observe that the Bayes factors for the \textsc{binned} models are always bounded from above by the Bayes factors for the \textsc{free} models. 

The model probabilities $P_n$ in Fig.~\ref{fig:bf} allow us to construct the BMA of the $\Omega$-PPL and $\gamma$-PPL bands in Fig.~\ref{fig:PPLnbands}. To this end, we combine at each frequency $f$ the $h^2\Omega_{\rm GW}$ and $\gamma$ distributions for the individual PPL models, weighting each model by its respective $P_n$ value, 
\begin{equation}
P\left(h^2\Omega_{\rm GW}|D\right) = \sum_{n=1}^6 P_n P\left(h^2\Omega_{\rm GW}|\textrm{PPL}n, D\right) \,,
\end{equation}
and similarly for the spectral index,
\begin{equation}
P\left(\gamma|D\right) = \sum_{n=1}^6 P_n P\left(\gamma|\textrm{PPL}n, D\right) \,.
\end{equation}
In both cases, we evaluate the model weights $P_n$ at their central values shown in Fig.~\ref{fig:bf}. The outcome of our analysis is shown in Figs.~\ref{fig:omegagammabands} and \ref{fig:og-binned-bma}. The BMA bands in these figures constitute the main result of this work. As in Fig.~\ref{fig:PPLnbands}, all bands in Figs.~\ref{fig:omegagammabands} and \ref{fig:og-binned-bma} indicate again the $68\%$ and $95\%$ median credible intervals of the corresponding posterior distributions at each frequency $f$. 

In view of our results, several comments are in order. First, it is apparent that the $\Omega$-PPL-BMA bands are dominated by their PPL1 contributions, in agreement with the fact that the model weight for PPL1 is the largest. Second, the BMA bands for the \textsc{binned} models reflect again the rigid frequency bins entering the construction of these models. At the same time, the results for the \textsc{binned} models serve as an approximation of the results for the \textsc{free} models. Finally, we conclude that the $\Omega$-PPL-BMA bands indeed provides us with a  physics-agnostic reconstruction of the signal in the NG15 data that assumes a middle role in between the simple CPL reconstruction and the more conservative free spectral reconstruction. First of all, our $\Omega$-PPL-BMA bands allow for the amount of complexity that is supported by the data. In particular, it is interesting to observe a slight upward trend around $f \simeq F_8 = \sfrac{8}{T_{\rm obs}}$, which is also visible in the NG15 violins, but which is not captured by the CPL reconstruction. While this feature is statistically not significant, it illustrates the local constraining power of our $\Omega$-PPL-BMA bands, which enable us to constrain local-in-frequency deviations from a globally smooth GWB spectrum. Second of all, our reconstruction of the NG15 signal is more continuous than the free spectral reconstruction. That is, it does not allow for large variations in GWB power from frequency $F_k$ to frequency $F_{k+1}$, which better reflects our physical expectation for a large class of spectral GWB models.

\begin{figure}
\begin{center}
\includegraphics[width=0.47\textwidth]{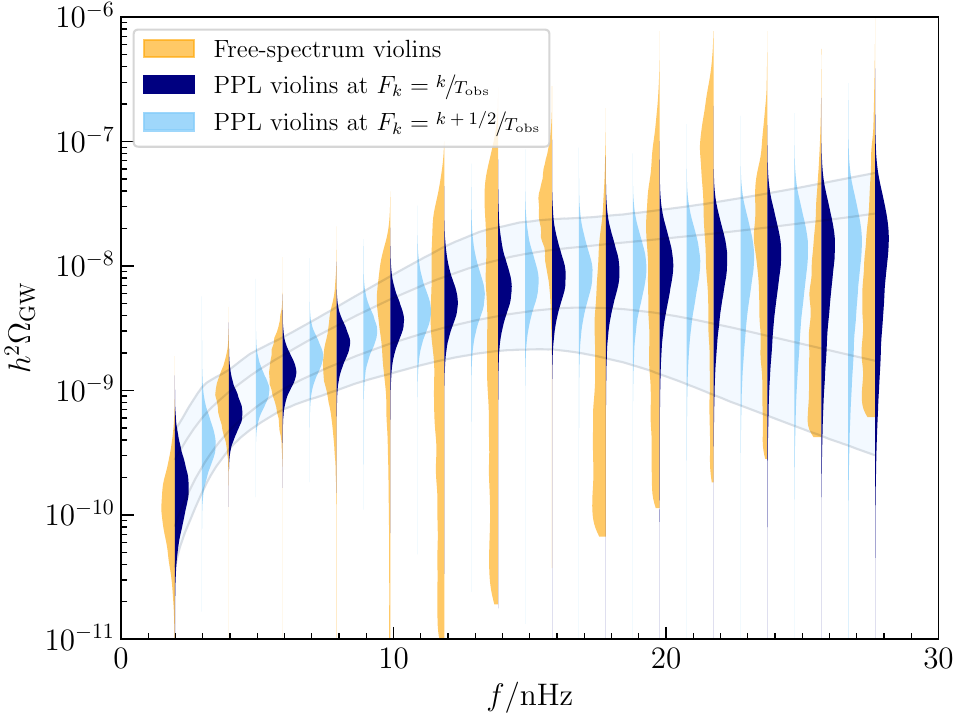}
\end{center}
\caption{Bayesian periodograms for the free spectral model and the BMA over the \textsc{PPL-free} models. The band in light blue in the background corresponds to the $\Omega$-PPL-BMA band for the \textsc{PPL-free} models; see the left panel of Fig.~\ref{fig:omegagammabands}.}
\label{fig:ppl-violins}
\end{figure}

The fact that our PPL reconstructions leads to tighter constraints on the GWB spectrum than the free spectral model can also be illustrated as follows. At each frequency $F_k = \sfrac{k}{T_{\rm obs}}$, we can extract the distribution of $h^2\Omega_{\rm GW}$ values from the $\Omega$-PPL-BMA band for the \textsc{free} models and use it to construct what we may refer to as a PPL violin. Together, these PPL violins result in the Bayesian periodogram for the BMA over our PPL models. We show this periodogram in Fig.~\ref{fig:ppl-violins} and compare it to the periodogram for the free spectral model, i.e., the standard NG15 violins. Clearly, the PPL violins follow the same trend as the free-spectrum violins. But at the same time, their extent along the $h^2\Omega_{\rm GW}$ axis is in most cases much shorter, reflecting their larger constraining power. We therefore conclude that, as expected, the NG15 data, in combination with the additional assumption of a PPL-type GWB spectrum, leads to stronger constraints on the GWB amplitudes than the more conservative free spectral model.

In particular, our  $\Omega$-PPL-BMA band for the \textsc{free} models allows us to construct an even larger number of PPL violins. Thanks to the assumption of a PPL-type GWB spectrum, our reconstruction of the NG15 signal provides us with $h^2\Omega_{\rm GW}$ posterior distributions at every $f$ between $f_{\rm min}$ and $f_{\rm max}$. To highlight this point, we show 13 additional PPL violins at half-integer multiples of $\sfrac{1}{T_{\rm obs}}$ in Fig.~\ref{fig:ppl-violins}, i.e., in between the violins at $F_k = \sfrac{k}{T_{\rm obs}}$. We emphasize that a similar construction is not possible in the free spectral model, which, by definition, only knows about the GWB amplitude at each $F_k$. In summary, this means that our PPL construction of the NG15 signal leads to a more constrained Bayesian periodogram with a finer (in principle, arbitrarily fine) frequency resolution than the free spectral model. 


\section{Outlook: PPL refits} 
\label{sec:application}


In the literature, numerous spectral GWB models have been proposed as possible interpretations of the NG15 signal. Standard parameter inference on these models requires a Bayesian MCMC fit to the NG15 timing residuals for each individual model, which is computationally expensive. Therefore, in order to facilitate faster model fits to the NG15 data, approximate refitting techniques have been developed in recent years.

The \texttt{ceffyl} framework~\citep{Lamb:2023jls}, e.g., builds upon the posterior densities for the amplitude parameters in the free spectral model (i.e., the NG15 violins). The basic idea behind the \texttt{ceffyl} framework is that the NG15 violins can be used to construct an induced likelihood on the parameter space of spectral GWB models. In a given spectral model, one simply evaluates the 14 GWB amplitudes $h^2\Omega_{\rm GW}(F_k)$ for each set of values of the underlying model parameters and reads off the posterior probabilities of these amplitudes from the 14 NG15 violins. The product of these 14 probabilities then defines, up to a constant normalization factor, the factorized \texttt{ceffyl} likelihood. The \texttt{ceffyl} refitting framework thus consists of two steps: first, one fits, once and for all, the free spectral model to the NG15 data; then, in a second step, one \textit{refits} other spectral GWB models to the amplitude posteriors obtained in the first step. 

A similar idea, but for a simpler physics-agnostic reference model, was recently proposed in~\cite{Esmyol:2025ket}, which discusses spectral refits to the RPL model. Again, this refitting framework consists of two steps: first, one fits, once and for all, the RPL model to the NG15 data; then, in a second step, one \textit{refits} other spectral GWB models to the posterior density obtained in the first step. Here, an advantage of the RPL model is the low dimensionality of its parameter space. While the free spectral model introduces a new amplitude parameter for each frequency bin, the RPL model features only three parameters: the amplitude $A$, the spectral index $\gamma$, and the running of the spectral index $\beta$. It is therefore easy to construct a numerical approximation of the full three-dimensional posterior density for $A$, $\gamma$, $\beta$, e.g., via kernel density estimation (KDE) applied to the MCMC chain that results from fitting the RPL model to the NG15 data. On the other hand, it is now less trivial to map arbitrary spectral GWB models onto the RPL parameter space, which is where the RPL posterior density lives. The analysis in \cite{Esmyol:2025ket} solves this problem by introducing a $\chi^2$-based matched-filtering approach that maps every input GWB spectrum to a triple of best-fit $A$, $\gamma$, and $\beta$ parameter values, i.e., a best-fit RPL proxy for the given input spectrum. This map notably allows one to pull back the RPL posterior density onto the parameter space of the spectral GWB model, where it can then be used as an induced likelihood.

The purpose of this short outlook section now is to point out that, next to refitting spectral models to the free spectral or RPL model, it would be interesting to refit spectral models to the PPL models discussed in the present work. As in the other two cases, the procedure would consist again of two steps:  first, one fits, once and for all, all relevant PPL model to the NG15 data; then, in a second step, one \textit{refits} other spectral GWB models to the PPL posterior densities obtained in the first step. The present paper provides a comprehensive completion of the first of these two tasks. In particular, the corner plots in Figs.~\ref{fig:corner-binned} and \ref{fig:corner-free} show (marginalized versions of) the PPL posterior densities that one would need for a PPL refitting analysis. As in~\cite{Esmyol:2025ket}, one could, e.g., attempt to construct KDE approximations of the respective PPL posterior densities, which live in $2n$ dimensions in the case of the \textsc{free} models.

In a second step, one would then need to construct a map from spectral GWB models to PPL-type spectra. We anticipate that $\chi^2$-based matched filtering might provide again a viable option. Furthermore, one would need to select not only best-fit points in the PPL parameter space, but also the most appropriate PPL model (CPL, BPL, etc.) to begin with. This additional model selection may also be accomplished by comparing the values of a $\chi^2$ statistic to each other. For now, however, we content ourselves with the fact that the posterior densities obtained in the present work provide the numerical basis for any PPL refit. We leave a discussion of all remaining details and the numerical implementation of a complete PPL refitting framework for future work. 

Finally, we conclude by stressing an important conceptual difference between \texttt{ceffyl}-style refits to the NG15 violins on the one hand and RPL or PPL refits on the other hand. In the case of refits to the free spectral model, the input GWB spectrum is always already contained in the class of spectra covered by the reference model (i.e., the free spectral model) by construction\,---\,at least under the assumptions of the current standard PTA analysis pipeline. The point is that, in the approach followed by \cite{NANOGrav:2023gor} and \cite{NANOGrav:2023hvm}, the map from the spectral model to the reference model is trivial: the GWB spectrum is simply modeled by a discrete Fourier series; information in between Fourier frequencies does not enter the analysis. The GWB spectrum is therefore exclusively represented by the set of amplitudes $h^2\Omega_{\rm GW}(F_k)$, which precisely correspond to the parameters of the free spectral model. The model-specific amplitudes $h^2\Omega_{\rm GW}(F_k)$ thus play the role of the coordinates of the input spectrum in the parameter space of the reference model. In the case of RPL or PPL refits, the situation is, by contrast, more complex. Most input spectra will not correspond precisely to a specific RPL or PPL spectrum. Hence, in these cases, it is necessary to first establish a map between models that can then be used to pull back the posterior density of the reference model onto the parameter space of the spectral model of interest. An important implication of this observation is that we cannot simply refit spectral models to the PPL violins. Instead, we first have to find the best-fit PPL proxy for a given input spectrum, before we can then evaluate the PPL posterior density at the location of this proxy spectrum. 


\section{Conclusions} 
\label{sec:conclusions}

In this paper, we presented a new physics-agnostic spectral reconstruction of the signal in the NANOGrav 15-year (NG15) data that is complementary to the well-known reconstructions based on the constant-power-law (CPL) model, running-power-law (RPL) model, or free spectral model. In our approach, we model the gravitational-wave background (GWB) spectrum in terms of a set of piecewise power laws (PPLs) and then marginalize over the number of PPLs in a Bayesian model average (BMA). The final result of this reconstruction is shown in Fig.~\ref{fig:omegagammabands} and can be regarded as the NG15 equivalent of other similar results in the literature, e.g., the PPL reconstruction of the primordial scalar power spectrum by the PLANCK collaboration based on the PLANCK 2015~\citep{Planck:2015sxf} and PLANCK 2018 data~\citep{Planck:2018jri}. Unlike the CPL reconstruction, our PPL reconstruction of the GWB spectrum accounts for the amount of complexity supported by the NG15 data. In particular, it is sensitive to local features in the GWB spectrum. Conversely, in the absence of local features, which is what we find for the NG15 data, our reconstruction allows us to constrain local-in-frequency deviations from a globally smooth GWB spectrum. At the same time, it offers a more continuous reconstruction of the GWB spectrum than the free spectrum, which better reflects our physical expectation, especially, from the perspective of cosmological sources. 

We also compared the PPL and free spectral reconstruction of the NG15 signal at the level of their Bayesian periodograms (see Fig.~\ref{fig:ppl-violins}), i.e., at the level of their amplitude violins, and commented on the possibility to use the PPL posterior densities shown in Figs.~\ref{fig:corner-binned} and \ref{fig:corner-free} in future PPL refit analyses. In particular, we discussed the ingredients that would be necessary to set up a complete PPL refitting framework\,---\,an interesting task that we leave for future work. Finally, we complemented every reconstruction of the GWB energy density power spectrum $h^2\Omega_{\rm GW}$ by an analogous reconstruction of the frequency-dependent spectral index $\gamma$. Our final result for the reconstruction of $\gamma$ is shown in Fig.~\ref{fig:omegagammabands}. When comparing the two panels in this figure, it is important to keep in mind that the panel for $\gamma$ does \textit{not} simply follow from taking the logarithmic frequency derivative of the $\Omega$-PPL-BMA band in the left panel. Instead, both the $\Omega$-PPL-BMA band in the left panel and the $\gamma$-PPL-BMA band in the right panel represent BMAs over distributions of $h^2\Omega_{\rm GW}$ and $\gamma$ values, at each frequency $f$, that follow from individual PPL spectra. 

In summary, we conclude that the PPL reconstruction of the NG15 signal represents a useful addition to the list of commonly employed reference models for the GWB, which motivates us to apply the methodology developed in this work also to future NANOGrav data sets. 


\section*{Supplementary information}

\smallskip
\centerline{\it Author contributions}
\medskip

This paper uses over a decade's worth of pulsar timing observations and is the product of the work of many people.
K.Sc.\ initiated and led the project, designed the analysis, and wrote the manuscript.
S.W. prepared first exploratory results.
A.B. and W.G.L. carried out the final numerical analysis.
W.G.L.\ prepared numerical data for cross-checks in the \texttt{ceffyl} framework.
B.D.G.C.\ and M.L.\ cross-checked parts of the numerical analysis based on \texttt{ceffyl}-type runs. 
W.G.L.\ and D.W.\ contributed to statistical and software development.
A.B.\ and K.Sc.\ prepared all figures.

\bigskip
\centerline{\it Acknowledgments}
\medskip

The authors would like to thank the Mainz Institute for Theoretical Physics (MITP) of the Cluster of Excellence PRISMA$+$ (Project ID 39083149) for its hospitality and support. The authors would also like to thank Joseph Lazio, Chiara Mingarelli, and Ken Olum for helpful feedback and comments on the manuscript as well as all members of the NANOGrav New Physics Working Group who provided feedback and comments. 

\bigskip
\centerline{\it Data set}
\medskip

G.A., A.A., A.M.A., Z.A., P.T.B., P.R.B., H.T.C., K.C., M.E.D., P.B.D., T.D., E.C.F., W.F., E.F., G.E.F., N.G.D., D.C.G., P.A.G., J.G., R.J.J., M.L.J., D.L.K., M.K., M.T.L., D.R.L., J.L., R.S.L., A.M., M.A.M., N.M., B.W.M., C.N., D.J.N., T.T.N., B.B.P.P., N.S.P., H.A.R., S.M.R., P.S.R., A.S., C.S., B.J.S.A., I.H.S., K.St., A.S., J.K.S., and H.M.W.\ developed timing models and ran observations for the NG15 data set.

\bigskip
\centerline{\it Computing resources}
\medskip

This work was conducted in part using the resources of the Advanced Computing Center for Research and Education (ACCRE) at Vanderbilt University, Nashville, TN.

\bigskip
\centerline{\it Funding information}
\medskip

The NANOGrav Collaboration receives support from National Science Foundation (NSF) Physics Frontiers Center award Nos.\ 1430284 and 2020265, the Gordon and Betty Moore Foundation, NSF AccelNet award No.\ 2114721, an NSERC Discovery Grant, and CIFAR. The Arecibo Observatory is a facility of the NSF operated under cooperative agreement (AST-1744119) by the University of Central Florida (UCF) in alliance with Universidad Ana G.\ M\'endez (UAGM) and Yang Enterprises (YEI), Inc. The Green Bank Observatory is a facility of the NSF operated under cooperative agreement by Associated Universities, Inc. The National Radio Astronomy Observatory is a facility of the NSF operated under cooperative agreement by Associated Universities, Inc. 

\smallskip
The work of A.B., B.D.G.C., M.L., and S.W.\ was supported in part by the U.S.\ Department of Energy under grant No.\ DE-SC0007914 and in part by Pitt PACC.
L.B.\ acknowledges support from the National Science Foundation under award AST-2307171 and from the National Aeronautics and Space Administration under award 80NSSC22K0808.
P.R.B.\ is supported by the Science and Technology Facilities Council, grant number ST/W000946/1.
S.B.\ gratefully acknowledges the support of a Sloan Fellowship, and the support of NSF under award \#1815664.
The work of R.B., R.C., X.S., J.T., and D.W.\ is partly supported by the George and Hannah Bolinger Memorial Fund in the College of Science at Oregon State University.
M.C.\ acknowledges support by the European Union (ERC, MMMonsters, 101117624).
Support for this work was provided by the NSF through the Grote Reber Fellowship Program administered by Associated Universities, Inc./National Radio Astronomy Observatory.
H.T.C.\ acknowledges funding from the U.S. Naval Research Laboratory.
Pulsar research at UBC is supported by an NSERC Discovery Grant and by CIFAR.
K.C.\ is supported by a UBC Four Year Fellowship (6456).
M.E.D.\ acknowledges support from the Naval Research Laboratory by NASA under contract S-15633Y.
T.D.\ and M.T.L.\ received support by an NSF Astronomy and Astrophysics Grant (AAG) award number 2009468 during this work.
E.C.F.\ is supported by NASA under award number 80GSFC24M0006.
G.E.F., S.C.S., and S.J.V.\ are supported by NSF award PHY-2011772.
K.A.G.\ and S.R.T.\ acknowledge support from an NSF CAREER award \#2146016.
D.C.G.\ is supported by NSF Astronomy and Astrophysics Grant (AAG) award \#2406919.
A.D.J.\ and M.V.\ acknowledge support from the Caltech and Jet Propulsion Laboratory President's and Director's Research and Development Fund.
A.D.J.\ acknowledges support from the Sloan Foundation.
N.La.\ was supported by the Vanderbilt Initiative in Data Intensive Astrophysics (VIDA) Fellowship.
Part of this research was carried out at the Jet Propulsion Laboratory, California Institute of Technology, under a contract with the National Aeronautics and Space Administration (80NM0018D0004).
W.G.L.\ was supported by a Royal Astronomical Society travel grant which inspired this work.
D.R.L.\ and M.A.M.\ are supported by NSF \#1458952.
M.A.M.\ is supported by NSF \#2009425.
C.M.F.M.\ was supported in part by the National Science Foundation under Grants No.\ NSF PHY-1748958 and AST-2106552.
A.Mi.\ is supported by the Deutsche Forschungsgemeinschaft under Germany's Excellence Strategy - EXC 2121 Quantum Universe - 390833306.
The Dunlap Institute is funded by an endowment established by the David Dunlap family and the University of Toronto.
K.D.O.\ was supported in part by NSF Grant No.\ 2207267.
T.T.P.\ acknowledges support from the Extragalactic Astrophysics Research Group at E\"{o}tv\"{o}s Lor\'{a}nd University, funded by the E\"{o}tv\"{o}s Lor\'{a}nd Research Network (ELKH), which was used during the development of this research.
P.P.\ and S.R.T.\ acknowledge support from NSF AST-2007993.
H.A.R.\ is supported by NSF Partnerships for Research and Education in Physics (PREP) award No.\ 2216793.
S.M.R.\ and I.H.S.\ are CIFAR Fellows.
Portions of this work performed at NRL were supported by ONR 6.1 basic research funding.
J.D.R.\ also acknowledges support from start-up funds from Texas Tech University.
K.Sc.\ is an affililate member of the Kavli Institute for the Physics and Mathematics of the Universe (Kavli IPMU) at the University of Tokyo and as such supported by the World Premier International Research Center Initiative (WPI), MEXT, Japan (Kavli IPMU).
J.S.\ is supported by an NSF Astronomy and Astrophysics Postdoctoral Fellowship under award AST-2202388, and acknowledges previous support by the NSF under award 1847938.
J.P.W.V.\ acknowledges support from NSF AccelNet award No.~2114721.
O.Y.\ is supported by the National Science Foundation Graduate Research Fellowship under Grant No.\ DGE-2139292.


\newpage
\bibliography{arxiv_1.bib}{}
\bibliographystyle{aasjournal}
\end{document}